\documentclass[final,5p,times,twocolumn]{elsarticle}
\usepackage{booktabs}
\usepackage[fleqn]{amsmath}
\usepackage{cases}
\usepackage{multirow}
\usepackage{xcolor}
\usepackage[normalem]{ulem}
\usepackage{tabularx}
\usepackage{siunitx}
\usepackage{comment}
\usepackage{algorithm}
\usepackage{algpseudocode}
\usepackage{algorithmicx}
\usepackage{grffile}
\usepackage{rotating}
\usepackage{lineno}
\usepackage{hyperref}
\usepackage{graphicx,enumerate,subfigure}
\usepackage{amssymb}
\usepackage{bm,amsmath}
\usepackage{caption,color,xcolor}
\usepackage{subeqnarray}
\usepackage{multirow}
\usepackage{lscape}
\usepackage{rotating}
\usepackage{graphics}
\usepackage{float}
\usepackage{epstopdf,epsfig}
\usepackage{threeparttable}
\usepackage{soul}
\usepackage{enumitem}

\usepackage[capitalize]{cleveref}

\floatname{algorithm}{Algorithm}
\biboptions{numbers,sort&compress} 
\journal{XXX}

\newcounter{bla}

\begin{document}
\nolinenumbers

\begin{frontmatter}
\title{An efficient treatment of heat-flux boundary conditions in  GSIS for rarefied gas flows}

 \author{Yanbing Zhang} 
 \author{Ruifeng Yuan} 
 \author{Liyan Luo} 
 \author{Lei Wu\corref{mycorrespondingauthor}}
 \cortext[mycorrespondingauthor]{Corresponding authors:}
 \ead{wul@sustech.edu.cn}

\address{Department of Mechanics and Aerospace Engineering, Southern University of Science and Technology, Shenzhen 518055, China}

\begin{abstract}
Heat-flux boundary conditions are challenging to implement efficiently in rarefied gas flow simulations because the wall-reflected gas temperature and density must be determined dynamically during the computation. This paper aims to tackle this problem within the general synthetic iterative scheme (GSIS), where the Boltzmann kinetic equation is solved deterministically in an outer loop and macroscopic synthetic equations are solved in an inner loop.
To avoid kinetic-macroscopic boundary-flux mismatch and the resulting convergence bottlenecks, for the macroscopic boundary flux at every inner iteration, the incident increment is estimated using a Maxwellian distribution, and then the reflected contribution is obtained by boundary conditions consistent with those in the kinetic solver.
In addition to retaining the fast-converging and asymptotic-preserving properties of GSIS, the proposed method significantly reduces the iterations required to determine the wall-reflected gas parameters. Numerical simulations of rarefied gas flows in and around a 3D nozzle, a 2D adiabatic cylinder, and a 2D annular heat-transfer configuration show good agreement with the direct simulation Monte Carlo method, while achieving substantial efficiency gains over conventional iterative schemes.
\end{abstract}


\end{frontmatter}

\section{Introduction}

Rarefied gas flows play an increasingly important role in modern engineering applications, such as spacecraft design, hypersonic flight, and vacuum technologies. In these situations, gas flows often depart from thermodynamic equilibrium, rendering the Navier-Stokes-Fourier (NSF) equations inadequate. The Boltzmann equation provides a fundamental mesoscopic description of gas dynamics and remains valid across all rarefaction regimes, making it essential for modeling rarefied gas flows. However, its numerical solution is highly challenging due to the high dimensionality of the velocity distribution function (VDF). As a result, solving the Boltzmann equation, through the stochastic direct simulation Monte Carlo (DSMC) method \cite{bird1994molecular} or the deterministic discrete velocity method \cite{Aristov2001Direct,mieussens2000}, incurs prohibitive computational cost. 

Over the past decades, a wide range of multiscale numerical methods have been developed to solve the Boltzmann equation and its simplified kinetic models. These include hybrid NSF-DSMC coupling approach \cite{sun2004hybrid}, unified gas-kinetic scheme and their variants \cite{Xu2010unified,guo2013discrete,LiuZhu2020JCP},  implicit-explicit Runge-Kutta method \cite{dimarcoPareschi2013,Dimarco2018JCP},
stochastic Fokker-Planck method~\cite{Gorji2014JCP}, 
general synthetic iterative scheme (GSIS)~\cite{su2020can}, and unified stochastic particle method \cite{Fei2020UnifiedBGK,Fei2023JCP}, among others.
Examining the development trends of these methods, it may be concluded that, to be efficient, they should satisfy several key criteria. First, \textbf{asymptotic-preserving (AP)}~\cite{klar1999}: in near-continuum flow regimes, conventional schemes that treat streaming and collision separately impose severe restrictions on spatial cell size and time step. An AP method, however, correctly recovers the NSF limit and allows the use of spatial cells much larger than the molecular mean free path~\cite{Xu2010unified,zhu2016implicit}. 
Second, \textbf{fast convergence}: since steady-state solutions are of primary interest in most applications, a numerical method should avoid unnecessary intermediate evolutions and converge rapidly to the steady state. 
Third, \textbf{universality}: because the Boltzmann collision operator becomes highly complex for polyatomic gases or chemically reacting flows, an effective numerical method should be applicable to a wide range of collision operators rather than being limited to simplified models such as the BGK or Shakhov models~\cite{Bhatnagar1954,shakhov1968}. Fourth, \textbf{simplicity and compatibility}: the numerical procedure should be as simple as possible and ideally compatible with conventional computational fluid dynamics frameworks, facilitating practical implementation and extension.

The GSIS has emerged as a representative framework that targets the above criteria~\cite{su2020can}. In GSIS, the steady-state solution is obtained by alternately solving a kinetic equation and a set of macroscopic synthetic equations on the same spatial grid. The synthetic equations are derived from the kinetic model, reduce to the NSF system in the near-continuum flow regime, and incorporate rarefaction effects through the velocity moments of the numerical VDF in the transition and free-molecular flow regimes.
The macroscopic synthetic solution, solved directly toward the steady state, is then fed back to guide the update of the VDF in the subsequent kinetic iteration. This two-way tight coupling endows the GSIS with fast-converging and AP properties \cite{su2020fast}. The GSIS is also compatible with finite-volume and discontinuous Galerkin discretizations, as well as established techniques in computational fluid dynamics. More recently, it has been extended from linearized to nonlinear kinetic equations, to polyatomic gas and gas-mixture models, to gas-radiation coupling \cite{zhu2021general,Zeng2023CaF,Zeng2025GSIS,Zeng2026GSIS}, and even to boost DSMC simulations \cite{Luo2024DIG,Luo2026DIG}, achieving computational efficiencies several orders of magnitude higher than those of conventional numerical schemes in large-scale 3D simulations \cite{zhang2024efficient}.

It should be noted that most multiscale methods in rarefied gas dynamics assume a fixed wall temperature. In experimentally relevant conditions for thermal-management components~\cite{Khan2021ConstHeatFlux}, heat-flux boundary conditions allow direct specification of energy transfer and yield consistent predictions across different Knudsen numbers. Unlike the continuum flow based on the NSF equations, implementing these conditions in kinetic simulation is challenging because the heat flux corresponds to a half-space moment of the VDF,  and for diffuse boundary condition the VDF reflected from the wall depends on unknown wall temperature and gas density, which must be determined by simultaneously enforcing both impermeability and the prescribed energy flux~\cite{Meng2015HeatFluxBC,Khan2021ConstHeatFlux}. Conventional numerical schemes, which lack global flow information exchange across the entire computational domain, require a prohibitively large number of iterations to determine the two unknown wall-reflected parameters. 
For example, in DSMC implementations of specified wall heat flux, the wall temperature is typically adjusted iteratively based on the sampled wall heat flux, and the convergence behavior depends on tunable relaxation/controlling factors and wall-sampling periods. As a result, millions of time steps are needed to determine the wall temperature in DSMC~\cite{wang2008heat,akhlaghi2012new,AkhlaghiRoohi2016DSMCHeatFlux}. 

Even in GSIS, where the synthetic equation is of diffusion type and thus enables efficient global exchange of flow information, the macroscopic boundary flux must remain compatible with the kinetic boundary condition and be updated consistently; failing to do so turns the near-wall region into a convergence bottleneck and can compromise robustness, especially for large implicit time steps. This paper is dedicated to developing a consistent and efficient macroscopic boundary treatment that preserves kinetic--macroscopic compatibility for prescribed thermal wall conditions within the GSIS framework.

The remainder of the paper is organized as follows. Section~\ref{sec:Governing_equations} presents the governing equations and summarizes the kinetic boundary-condition settings. Section~\ref{sec:gsis} describes the numerical methodology, including the GSIS algorithm and the proposed consistent macroscopic boundary treatment used in the synthetic equations. Section~\ref{sec:num_example} reports the computational results and provides detailed discussion on accuracy, robustness, convergence behavior, and efficiency. Finally, conclusions are given in Section~\ref{sec:conclusion}.

\section{Governing equations}\label{sec:Governing_equations}

In gas kinetic theory, the VDF $f(t,\bm{x},\bm{v})$ describes the state of a monatomic gas with three translational degrees of freedom, where $t$ is time, $\bm{x}=(x_1,x_2,x_3)$ is the spatial coordinate, and $\bm{v}=(v_1,v_2,v_3)$ is the molecular velocity. 
The macroscopic variables are obtained by taking velocity moments of the VDF. For examples, the density $\rho$, flow velocity $\bm{u}$, and total energy per unit mass $E$ are defined as 
\begin{equation}\label{eq:moment_mono}
\left(\rho,~\rho\bm{u},~\rho E\right)=\int_{\mathbb{R}^3}\left(1,~\bm{v},~\frac{1}{2}v^2\right)f\,\mathrm{d}\bm{v}.
\end{equation}
The specific internal energy is $e=E-u^2/2=\frac{3}{2}RT$, and the pressure is $p=\rho RT$, where $R$ is the specific gas constant. The deviatoric stress tensor $\bm{\sigma}$ and the heat flux $\bm{q}$ are given by
\begin{equation}\label{eq:stress_q_mono}
\bm{\sigma}=\int_{\mathbb{R}^3}\left(\bm{c}\bm{c}-\frac{c^2}{3}\mathrm{I}\right)f\,\mathrm{d}\bm{v},\quad
\bm{q}=\int_{\mathbb{R}^3}\frac{1}{2}c^2\,\bm{c}\,f\,\mathrm{d}\bm{v},
\end{equation}
where $\mathrm{I}$ is the $3\times 3$ identity matrix, and $\bm{c}=\bm{v}-\bm{u}$ is the  peculiar velocity. 

\begin{table*}[!t]
\centering
\caption{Summary of boundary specifications and unknown wall parameters.}\label{tab:bc_summary}
\begin{tabular}{lccc}
\hline
Wall type & Prescribed data & Unknowns & Discrete closure conditions\\
\hline
isothermal & $T_w=T_0$ & $\rho_w$ & zero net mass flux\\
adiabatic  & $Q_n=0$ & $(T_w,\rho_w)$ & zero net mass flux, zero net energy flux\\
heat-flux & $Q_n\neq 0$ & $(T_w,\rho_w)$ & zero net mass flux, prescribed net energy flux\\
far-field & $(\rho_\infty,\bm{u}_\infty,T_\infty)$ & -- & incoming Maxwellian\\
\hline
\end{tabular}
\end{table*}

\subsection{The Shakhov gas-kinetic model}

Without loss of generality, we assume the evolution of VDF is governed by the Shakhov kinetic model~\cite{shakhov1968}:
\begin{equation}\label{eq:shakhov_model}
\frac{\partial f}{\partial t}+\bm{v}\cdot\nabla f=\frac{g_S-f}{\tau},
\end{equation}
where $\tau=\mu/p$ is the mean collision time of gas molecules, and $\mu$ is the shear viscosity taking the power-law form as
\begin{equation}\label{eq:viscosity_powerlaw}
\mu(T)=\mu(T_0)\left(\frac{T}{T_0}\right)^{\omega},
\end{equation}
with $\omega$ the viscosity index and $T_0$ the reference temperature. The thermal conductivity is given by $\kappa=15\mu R/4$, where the Prandtl number is taken to be 2/3 for a monatomic gas.

The reference VDF $g_S$ is constructed in the following manner to recover the shear viscosity and thermal conductivity in the continuum limit:
\begin{equation}\label{eq:shakhov_gS}
g_S=g_M\left[1+\frac{\bm{c}\cdot\bm{q}}{15pRT}\left(\frac{c^2}{RT}-5\right)\right],
\end{equation}
where $g_M$ is the local Maxwellian
\begin{equation}\label{eq:maxwellian}
g_M=\rho\left(\frac{1}{2\pi RT}\right)^{3/2}\exp\left(-\frac{c^2}{2RT}\right).
\end{equation}

Taking velocity moments of Eq.~\eqref{eq:shakhov_model} yields the macroscopic conservation laws for mass, momentum, and total energy:
\begin{equation}\label{eq:macro_mono}
\begin{aligned}
\frac{\partial\rho}{\partial t}+\nabla\cdot(\rho\bm{u})&=0,\\
\frac{\partial(\rho\bm{u})}{\partial t}+\nabla\cdot(\rho\bm{u}\bm{u})+\nabla\cdot\bm{P}&=0,\\
\frac{\partial(\rho E)}{\partial t}+\nabla\cdot(\rho E\bm{u})+\nabla\cdot(\bm{P}\cdot\bm{u}+\bm{q})&=0,
\end{aligned}
\end{equation}
where the pressure tensor is $\bm{P}=p\,\mathrm{I}+\bm{\sigma}$. 

In the continuum regime, i.e., when the Knudsen number
\begin{equation}\label{eq:Kn_def}
\mathrm{Kn}=\frac{\lambda}{L}\equiv\frac{\mu(T_0)}{p_0 L}\sqrt{\frac{\pi R T_0}{2}}
\end{equation}
is small, performing a Chapman--Enskog expansion on Eq.~\eqref{eq:shakhov_model} recovers the NSF constitutive relations \cite{chapman1990mathematical}:
\begin{equation}\label{eq:NSF_mono}
\begin{aligned}
&\bm{\sigma}_\text{NSF}=-\mu\left(\nabla\bm{u}+\nabla\bm{u}^{\mathrm{T}}-\frac{2}{3}(\nabla\cdot\bm{u})\mathrm{I}\right),\\
&
\bm{q}_\text{NSF}=-\kappa\nabla T.
\end{aligned}
\end{equation}

\subsection{Kinetic boundary conditions}\label{subsec:bc}

Let $\bm{n}$ denote the outward unit normal pointing from the gas domain to the boundary. We split the molecular velocity space into the discrete nodes $\Xi^+=\{\bm{v}_k:\bm{v}_k\cdot\bm{n}>0\}$ (incident molecules, outgoing from the gas to the boundary) and $\Xi^-=\{\bm{v}_k:\bm{v}_k\cdot\bm{n}\le 0\}$ (reflected molecules, incoming from the boundary to the gas), with $\Delta\Xi$ being the velocity-space discretization weight used in calculating the moments of the VDF. For all boundaries, the outgoing part $f_b(\bm{v}_k)$ for $\bm{v}_k\in\Xi^+$ is taken from the interior kinetic reconstruction, while the incoming part $f_b(\bm{v}_k)$ for $\bm{v}_k\in\Xi^-$ is prescribed by the corresponding boundary model.


For a stationary wall with diffuse reflection, the incoming distribution is modeled by a wall Maxwellian
\begin{equation}\label{eq:bc_wall_maxwellian}
\begin{aligned}
&f_b(\bm{v}_k)=\rho_w\,g_{w,k}(T_w),\\
&g_{w,k}(T)=\left(\frac{1}{2\pi RT}\right)^{3/2}\exp\!\left(-\frac{v_k^2}{2RT}\right),\quad \bm{v}_k\in\Xi^-.
\end{aligned}
\end{equation}
where $\rho_w$ and $T_w$ are determined by wall constraints. In practice, we enforce these constraints in a fully discrete form to eliminate quadrature mismatch between the kinetic boundary and the macroscopic moments.

For clarity and to avoid repeating similar derivations, Table~\ref{tab:bc_summary} summarizes all boundary conditions used in this work in terms of (i) the prescribed macroscopic data, (ii) the unknown wall parameters to be determined in the diffuse-reflection model \eqref{eq:bc_wall_maxwellian}, and (iii) the discrete constraints enforced at each boundary face. In particular, the three wall boundaries (isothermal, adiabatic, and heat-flux) share the same incoming Maxwellian form \eqref{eq:bc_wall_maxwellian}; their difference lies only in whether $T_w$ is prescribed and in the energy-flux constraint (zero for adiabatic wall and prescribed $Q_n$ for heat-flux wall). The far-field boundary is treated as an open boundary by prescribing an incoming Maxwellian corresponding to $(\rho_\infty,\bm{u}_\infty,T_\infty)$, while taking the outgoing part from the interior reconstruction. The following subsections provide the detailed discrete formulations and the procedures for computing the wall parameters.

\subsubsection{Isothermal wall}\label{subsubsec:bc_isothermal}

For an isothermal wall, the wall temperature is prescribed as $T_w=T_0$, and only $\rho_w$ needs to be determined to satisfy the discrete impermeability (zero net mass flux) condition. Define the outgoing mass flux moment from the interior as
\begin{equation}\label{eq:bc_Mplus}
M^+ = \sum_{\bm{v}_k\in\Xi^+} (\bm{v}_k\!\cdot\!\bm{n})\,f_k\,\Delta\Xi_k,
\end{equation}
and the incoming unit-density Maxwellian half-moment as
\begin{equation}\label{eq:bc_A_of_T}
A(T) = \sum_{\bm{v}_k\in\Xi^-} (\bm{v}_k\!\cdot\!\bm{n})\,g_{w,k}(T)\,\Delta\Xi_k.
\end{equation}
The discrete no-penetration constraint reads $M^+ + \rho_w A(T_0)=0$, hence
\begin{equation}\label{eq:bc_rhow_isothermal}
\rho_w = -\frac{M^+}{A(T_0)}.
\end{equation}

\subsubsection{Adiabatic wall and heat-flux wall}

For an adiabatic wall or a prescribed heat-flux wall, the net normal energy flux across the wall is specified, and both $T_w$ and $\rho_w$ are unknown under the diffuse-reflection model.
In addition to the mass moment $M^+$ in Eq.~\eqref{eq:bc_Mplus}, we define the outgoing energy flux moment from the interior
\begin{equation}\label{eq:bc_Qplus}
Q^+ = \sum_{\bm{v}_k\in\Xi^+} \frac{1}{2}m v_k^2\,(\bm{v}_k\!\cdot\!\bm{n})\,f_k\,\Delta\Xi_k,
\end{equation}
and the incoming unit-density Maxwellian energy half-moment
\begin{equation}\label{eq:bc_B_of_T}
B(T) = \sum_{\bm{v}_k\in\Xi^-} \frac{1}{2}m v_k^2\,(\bm{v}_k\!\cdot\!\bm{n})\,g_{w,k}(T)\,\Delta\Xi_k.
\end{equation}
Together with the incoming unit-density Maxwellian mass half-moment $A(T)$ in Eq.~\eqref{eq:bc_A_of_T}, the discrete impermeability condition gives $\rho_w(T)=-M^+/A(T)$.
We adopt the sign convention that $Q_n>0$ corresponds to heating the gas (i.e., energy entering the gas from the wall).
Since $\bm{n}$ points outward from the gas to the wall, the discrete energy balance reads
\begin{equation}\label{eq:bc_heatflux_energy}
Q^+ + \rho_w(T_w)\,B(T_w) = -\,Q_n,
\end{equation}
where $Q_n=0$ for an adiabatic wall and $Q_n\neq 0$ for a prescribed heat-flux wall.
Substituting $\rho_w(T)=-M^+/A(T)$ yields the scalar nonlinear equation
\begin{equation}\label{eq:bc_R_heatflux}
\begin{aligned}
\mathcal{R}(T_w) = Q^+ + Q_n - \frac{M^+\,B(T_w)}{A(T_w)} = 0.
\end{aligned}
\end{equation}

We employ the Newton iteration to solve for $T_w$.
Starting from the initial guess $T^{(0)}=(Q^+ + Q_n)/(2RM^+)$, we perform the update
\[
T^{(\ell+1)} = T^{(\ell)}-\frac{\mathcal{R}\!\left(T^{(\ell)}\right)}{\mathcal{R}'\!\left(T^{(\ell)}\right)},
\]
where
\begin{equation}\label{eq:bc_Rprime_general}
\begin{aligned}
\mathcal{R}'(T) &= -M^+\,\frac{B'(T)A(T)-B(T)A'(T)}{A(T)^2},\\
A'(T)&=\sum_{\bm{v}_k\in\Xi^-} (\bm{v}_k\!\cdot\!\bm{n})\,\frac{\partial g_{w,k}}{\partial T}\,\Delta\Xi_k,\\
B'(T)&=\sum_{\bm{v}_k\in\Xi^-} \frac{1}{2}m v_k^2(\bm{v}_k\!\cdot\!\bm{n})\,\frac{\partial g_{w,k}}{\partial T}\,\Delta\Xi_k,
\end{aligned}
\end{equation}
with
\begin{equation}\label{eq:bc_dg_dT}
\frac{\partial g_{w,k}}{\partial T}=\left(-\frac{3}{2T}+\frac{v_k^2}{2RT^2}\right)g_{w,k}.
\end{equation}

Upon convergence, we set $T_w=T^{(\ell_\star)}$ and compute the wall density as $\rho_w=-M^+/A(T_w)$.

\subsubsection{Far-field boundary}\label{subsubsec:bc_farfield}

At the far-field boundary, a macroscopic state $(\rho_\infty,\bm{u}_\infty,T_\infty)$ is prescribed according to the test case. The incoming distribution for $\bm{v}_k\in\Xi^-$ is set to the corresponding Maxwellian
\begin{equation}\label{eq:bc_farfield_maxwellian}
f_b(\bm{v}_k)=\rho_\infty\left(\frac{1}{2\pi RT_\infty}\right)^{3/2}\exp\!\left(-\frac{|\bm{v}_k-\bm{u}_\infty|^2}{2RT_\infty}\right),\quad \bm{v}_k\in\Xi^-.
\end{equation}

\section{Numerical methods}\label{sec:gsis}

We adopt the finite-volume scheme with second-order accuracy to solve the kinetic equations and synthetic equations.
Here we only briefly outline the main steps and refer the reader to Refs.~\cite{zhang2024efficient, liu2024further} for details of GSIS; the implementation of the heat-flux and related boundary conditions is elaborated in Section~\ref{subsec:macro_bc}.


\subsection{CIS method}

Conventional iterative schemes (CIS) update the kinetic equation alone. For each discrete velocity $\bm{v}_k\in\Xi$, the finite-volume discretization of the Shakhov model \eqref{eq:shakhov_model} reads
\begin{equation}
\label{eq:cis}
\frac{f^{\,n+1}_{i,k}-f^{\,n}_{i,k}}{\Delta t}+\frac{1}{V_i}\sum_{j\in N(i)}(\bm{v}_k\!\cdot\!\bm{n}_{ij})\,f^{\,n+1}_{ij,k}\,S_{ij}=\frac{g^{\,n}_{S,i,k}-f^{\,n+1}_{i,k}}{\tau^{\,n}_i},
\end{equation}
where $\Delta t$ is the pseudo-time step for steady iteration, $V_i$ is the cell volume, $S_{ij}$ is the face area, and $\bm{n}_{ij}$ is the unit normal pointing from cell $i$ to its neighbor $j$. The interface distribution $f_{ij,k}^{\,n+1}$ is reconstructed by a second-order upwind scheme based on the sign of $\bm{v}_k\!\cdot\!\bm{n}_{ij}$ (and the same half-space prescription is used for boundary faces). The relaxation time is $\tau_i=\mu(T_i)/p_i$.

\subsection{GSIS method}\label{subsec:GSIS}

\begin{figure}[!t]
    \centering
    {\includegraphics[width=0.49\textwidth,clip = true]{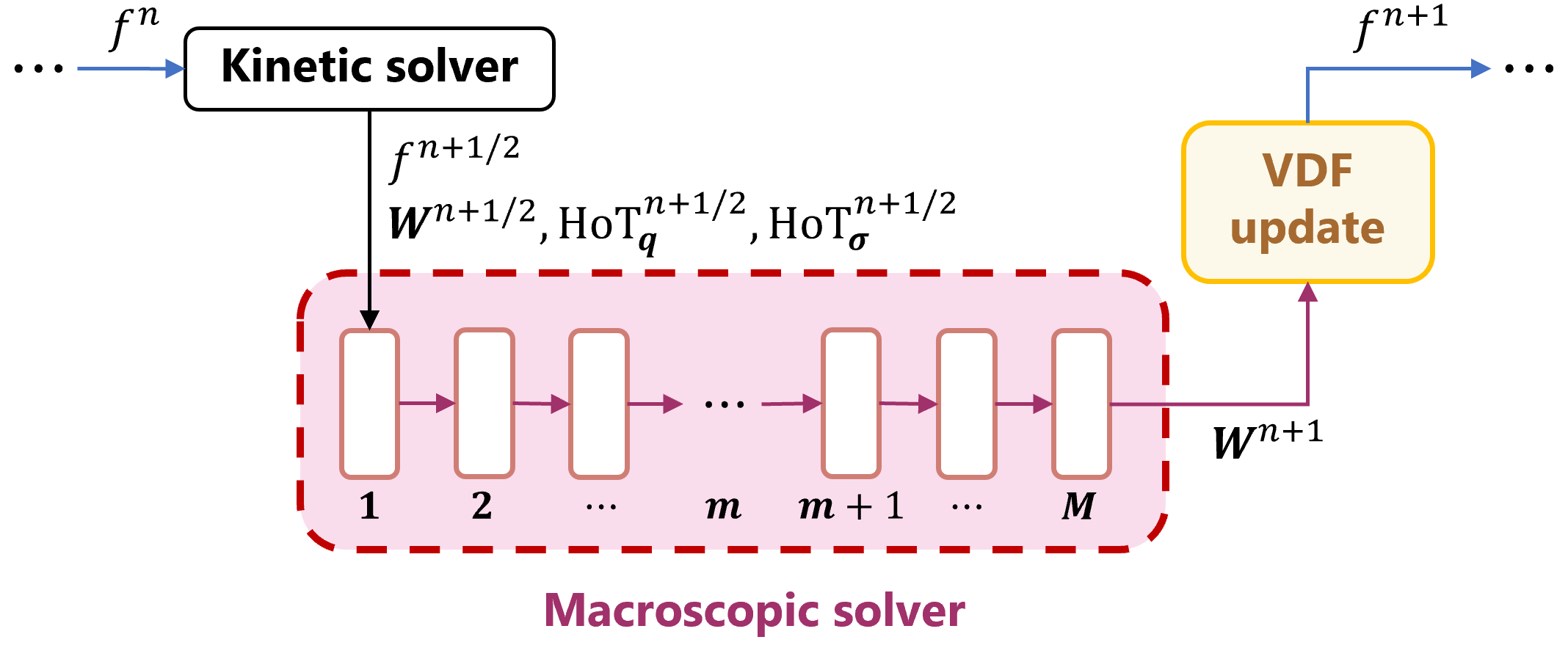}}
    \caption{Schematic for the GSIS algorithm: the kinetic solver computes an intermediate distribution $f^{n+1/2}$ using Eq.~\eqref{eq:cis} and the associated macroscopic quantities, followed by $M$ macroscopic inner iterations to solve Eq.~\eqref{eq:macro_update_mono}, where $m$ denotes the $m$-th inner macroscopic iteration, and a final VDF update as per Eq.~\eqref{eq:updatef_mono} to obtain $f^{n+1}$.}
    \label{fig:GSIS_Process}
\end{figure}

One GSIS iteration consists of one kinetic update followed by multiple inner iterations of the macroscopic synthetic system. Given the solution at the $n$-th GSIS step, the kinetic update produces an intermediate state denoted by $n+1/2$, and the macroscopic inner iterations subsequently drive the conservative variables to the next GSIS step $n+1$. The detailed algorithmic flow is illustrated in Figure~\ref{fig:GSIS_Process}.

For each discrete molecular velocity $\bm{v}_k\in\Xi$, the finite-volume discretization of the Shakhov model reads
\begin{equation}\label{eq:dvm_discrete_shakhov}
\frac{f^{\,n+1/2}_{i,k}-f^{\,n}_{i,k}}{\Delta t}+\frac{1}{V_i}\sum_{j\in N(i)}(\bm{v}_k\!\cdot\!\bm{n}_{ij})\,f^{\,n+1/2}_{ij,k}\,S_{ij}=\frac{g^{\,n}_{S,i,k}-f^{\,n+1/2}_{i,k}}{\tau^{\,n}_i}.
\end{equation}
Then, the VDF in GSIS is corrected so that its equilibrium part matches the updated macroscopic state:
\begin{equation}\label{eq:updatef_mono}
f^{n+1}_{i,k}=f^{n+1/2}_{i,k}+\left[f_{eq}\!\left(\bm{W}^{n+1}_i\right)-f_{eq}\!\left(\bm{W}^{n+1/2}_i\right)\right]_k,
\end{equation}
where $\bm{W}=(\rho,\rho\bm{u},\rho E)^{\top}$ are the conservative variables and $f_{eq}(\bm{W})\equiv g_M(\rho,\bm{u},T)$ is the Maxwellian corresponding to $\bm{W}$.

The macroscopic variables are obtained by solving a synthetic system that takes the NSF equations as the leading-order closure while incorporating high-order rarefaction effects through source terms extracted from $f^{n+1/2}$. Specifically, the stress and heat flux are written as
\begin{equation}\label{eq:full_constitutive_mono}
    \begin{aligned}
        \bm{\sigma}^{n+1}&=\bm{\sigma}^{n+1}_\text{NSF}+\underbrace{\int_{\mathbb{R}^3}\left(\bm{c}\bm{c}-\frac{c^2}{3}\mathrm{I}\right)f^{n+1/2}\,\mathrm{d}\bm{v}-\bm{\sigma}^{n+1/2}_\text{NSF}}_{\mathrm{HoT}_{\bm{\sigma}}}, \\
\bm{q}^{\,n+1}&=\bm{q}^{n+1}_\text{NSF}+\underbrace{\frac12\int_{\mathbb{R}^3}c^2\bm{c}\,f^{n+1/2}\,\mathrm{d}\bm{v}-\bm{q}^{n+1/2}_\text{NSF}}_{\mathrm{HoT}_{\bm{q}}}.
    \end{aligned}
\end{equation}
Substituting Eq.~\eqref{eq:full_constitutive_mono} into Eq.~\eqref{eq:macro_mono} yields an NSF-type system with rarefaction source terms, which is solved by an implicit finite-volume scheme to update $\bm{W}_i$:
\begin{equation}\label{eq:macro_update_mono}
\frac{\bm{W}_i^{\,n+1}-\bm{W}_i^{\,n+1/2}}{\Delta t}+\frac{1}{V_i}\sum_{j\in N(i)}\bm{F}^{\,n+1}_{ij}\,S_{ij}=\bm{0},
\end{equation}
where the macroscopic numerical flux $\bm{F}_{ij}$ includes both convective and viscous contributions and is evaluated in accordance with the constitutive relations in Eq.~\eqref{eq:full_constitutive_mono}. In particular, for interior faces the convective flux is computed by the low-dissipation and high-robustness SLAU scheme~\cite{shima2011slau}, while the viscous contribution is discretized consistently with Eq.~\eqref{eq:full_constitutive_mono}. 
At boundary faces, however, directly applying the same macroscopic flux construction may neglect rarefaction effects, leading to incorrect results; 
therefore, the boundary flux required by Eq.~\eqref{eq:macro_update_mono} is supplied by a treatment that remains fully consistent with the kinetic boundary condition at the discrete level. After the macroscopic inner iterations converge (or after a specified number of inner iterations), $\bm{W}^{n+1}$ is obtained, and the kinetic distribution is subsequently updated to $f^{n+1}$ using Eq.~\eqref{eq:updatef_mono}, thereby completing one GSIS step. We next describe this consistent kinetic--macroscopic boundary-flux treatment for the synthetic equations.



\subsection{Consistent kinetic-macroscopic boundary flux}\label{subsec:macro_bc}

Since rarefied gas flows exhibit velocity slip and temperature jump at the solid boundary, the macroscopic boundary treatment in the synthetic equations is nontrivial and critically affects overall efficiency and robustness of GSIS. While the kinetic boundary condition prescribes the boundary VDF in a half-space sense (incoming versus outgoing velocities), the macroscopic synthetic equations are solved in an inner loop to accelerate convergence; therefore, the macroscopic boundary flux must remain compatible with the kinetic boundary condition and, crucially, must be updated consistently during the macroscopic inner iterations. Otherwise, the boundary region may become the dominant bottleneck for fast convergence and can even trigger instability when large implicit steps are used in the macroscopic solver.


Early macroscopic boundary treatment in GSIS excluded near-wall layers from the synthetic subdomain~\cite{zhu2021general}, effectively freezing boundary contributions and impeding the propagation of kinetic non-equilibrium information. To mitigate this issue, subsequent incremental-feedback approaches updated boundary fluxes using macroscopic increments~\cite{Zeng2023CaF}; however, they remained indirect and failed to enforce kinetic constraints explicitly. More recently, the generalized boundary treatment reconstructs a truncated Grad-type distribution within each macroscopic inner iteration from conserved variables plus kinetic non-equilibrium corrections~\cite{liu2024further}. 
Despite these advances, it exhibits two limitations: (i) systematic flux inconsistencies arising from mismatched integration operators (continuous moments vs. discrete quadratures); and (ii) deteriorating fidelity and robustness at high Knudsen numbers, where truncated moment reconstructions fail to adequately capture strongly non-Maxwellian, anisotropic near-wall VDFs, leading to oscillations, nonphysical behavior, and numerical instabilities inherited from moment methods~\cite{thatcher2008g13bc,torrilhon2008r13bc}.
These observations underscore the need for a boundary-flux update that remains consistent with the kinetic boundary condition throughout the macroscopic inner iterations, which we detail next.

As shown in Fig.~\ref{fig:GSIS_Process}, one GSIS step consists of two stages: (i) a kinetic update producing an intermediate distribution $f^{n+1/2}$, and (ii) $M$ inner iterations of a macroscopic synthetic system that drives the conservative variables from $\bm W^{n+1/2,0}$ to $\bm W^{n+1/2,M}\equiv \bm W^{n+1}$. During the macroscopic inner iterations, the kinetic distribution is not recomputed; instead, the macroscopic boundary flux is updated in a way that remains consistent with the kinetic boundary condition while reflecting the evolving macroscopic state.

\subsubsection{Half-space splitting and continuous flux moments}\label{subsubsec:macro_bc_defs}


In practice, the boundary flux of conservative variables is split into outgoing and incoming contributions:
\begin{equation}\label{eq:macro_flux_cont_def}
\begin{aligned}
\bm F_b =
\underbrace{\int_{\Xi^+} (\bm v\!\cdot\!\bm n)\, f_b\,\bm\psi\,\mathrm d\bm v}_{\bm F_b^{\mathrm{out}}}
+\underbrace{\int_{\Xi^-} (\bm v\!\cdot\!\bm n)\, f_b\,\bm\psi\,\mathrm d\bm v}_{\bm F_b^{\mathrm{in}}}
,
\end{aligned}
\end{equation}
where $\bm\psi=(1,\bm v,\tfrac12 v^2)^\top$.


While the macroscopic solver is written in global Cartesian coordinates, it is better to enforce boundary conditions in geometry-aligned normal--tangential components. We introduce a local orthonormal frame $(\bm n,\bm t_1,\bm t_2)$ and local velocity components $(v_n,v_{t1},v_{t2})$. We also  denote the local-coordinate conservative state by $\bm W_{\mathrm{loc}}=(\rho,\rho u_n,\rho u_{t1},\rho u_{t2},\rho E)^\top$. The mapping between global and local conservative variables is written as
\begin{equation}\label{eq:macro_Q_matrix}
\bm W=\mathbf P {} \bm W_{\mathrm{loc}},
\quad \text{with}~~
\mathbf P=
\left(\begin{array}{ccccc}
1 & 0 & 0 & 0 & 0\\
0 & n_x & t_{1x} & t_{2x} & 0\\
0 & n_y & t_{1y} & t_{2y} & 0\\
0 & n_z & t_{1z} & t_{2z} & 0\\
0 & 0 & 0 & 0 & 1
\end{array}\right).
\end{equation}
and the corresponding flux vectors are transformed by $\bm F=\mathbf P\,\bm F_{\mathrm{loc}}$.


During macroscopic iterations, we approximate the change of the outgoing flux by the change of the equilibrium half-range moment. Let $f_{eq}(\bm W)$ be the Maxwellian associated with the local macroscopic state $\bm W$ (with velocity $\bm u$ and temperature $T$). We define the outgoing equilibrium half-range moment in the local frame as
\begin{equation}\label{eq:Floc_plus_def}
\bm F_{\mathrm{loc}}^{+}(\bm W_{\mathrm{loc}}) = \int_{v_n>0} v_n\, f_{eq}(\bm W_{\mathrm{loc}})\,\bm\psi_{\mathrm{loc}}\,\mathrm d\bm v,
\end{equation}
where $\bm\psi_{\mathrm{loc}}=(1,v_n,v_{t1},v_{t2},\tfrac12 v^2)^\top$.

For a stationary diffuse wall, the incoming (reflected) distribution is a wall Maxwellian $f_b=\rho_w g_w(T_w)$ on $\Xi^-$, where $g_w(T)=(2\pi RT)^{-3/2}\exp(-v^2/2RT)$. We define the unit-density incoming half-range moment as
\begin{equation}\label{eq:Fminus_hat_def}
\widehat{\bm F}_{\mathrm{loc}}^{-}(T_w) = \int_{v_n\le 0} v_n\, g_w(T_w)\,\bm\psi_{\mathrm{loc}}\,\mathrm d\bm v,
\end{equation}
so that the incoming macroscopic boundary flux is $\bm F_{\mathrm{loc}}^{-}(T_w,\rho_w)=\rho_w\,\widehat{\bm F}_{\mathrm{loc}}^{-}(T_w)$.

Let $a=\sqrt{RT}$, $u_t^2=u_{t1}^2+u_{t2}^2$, $\xi=u_n/\sqrt{2}a$, $\Phi=1-\mathrm{erfc}(\xi)/2$, and $\Psi=\exp(-\xi^2)/\sqrt{2\pi}$. Then the outgoing equilibrium half-range moment obtained by the continuous velocity-space integration takes the form
\begin{equation}\label{eq:Floc_plus_closed}
\bm F_{\mathrm{loc}}^{+}(\bm W_{\mathrm{loc}})=\rho
\begin{bmatrix}
H_1\\
H_2\\
u_{t1}H_1\\
u_{t2}H_1\\
\tfrac12\big(H_3+H_1(u_t^2+2a^2)\big)
\end{bmatrix},
\end{equation}
with
\begin{equation}\label{eq:H123_closed}
\begin{aligned}
& H_1=u_n\Phi+a\Psi,\\
& H_2=\left(u_n^2+a^2\right)\Phi+u_n a\Psi,\\
& H_3=\left(u_n^3+3u_n a^2\right)\Phi+a\left(u_n^2+2a^2\right)\Psi.
\end{aligned}
\end{equation}
For the stationary wall Maxwellian, the unit-density incoming half-range moment admits the closed form
\begin{equation}\label{eq:Fminus_hat_closed}
\widehat{\bm F}_{\mathrm{loc}}^{-}(T_w)=
\begin{bmatrix}
-\sqrt{\dfrac{RT_w}{2\pi}}\\[4pt]
\dfrac12 RT_w\\[4pt]
0\\[2pt]
0\\[4pt]
-\sqrt{\dfrac{2RT_w}{\pi}}\,RT_w
\end{bmatrix}.
\end{equation}

\subsubsection{Initialization after the kinetic iteration stage}\label{subsubsec:macro_bc_init}

After solving the kinetic equation, we obtain $f^{n+1/2}$ and assemble the boundary distribution $f_b^{n+1/2}$ using the same kinetic boundary condition as in Section~\ref{subsec:bc}. The macroscopic boundary flux at the start of the macroscopic stage is recorded by the discrete integration in the discretized velocity space
\begin{equation}\label{eq:macro_flux_proj}
\bm F_b^{\,m=0}=\sum_k (\bm v_k\!\cdot\!\bm n)\, f_{b,k}^{\,n+1/2}\,\bm\psi_k\,\Delta\Xi_k,
\end{equation}
where $m$ denotes the inner macroscopic iteration index, as shown in Figure~\ref{fig:GSIS_Process}. Note that for brevity we omit the superscript $n+1/2$ denoting the outer kinetic iteration step for the macroscopic boundary flux. Accordingly, the outgoing part is stored as
\begin{equation}\label{eq:macro_flux_outgoing}
\bm F_b^{\mathrm{out},m=0}=\sum_{\bm v_k\in\Xi^+}(\bm v_k\!\cdot\!\bm n)\, f_{b,k}^{\,n+1/2}\,\bm\psi_k\,\Delta\Xi_k,
\end{equation}
and the corresponding incoming part is stored as
\begin{equation}\label{eq:macro_flux_incoming}
\bm F_b^{\mathrm{in},m=0}=\sum_{\bm v_k\in\Xi^-}(\bm v_k\!\cdot\!\bm n)\, f_{b,k}^{\,n+1/2}\,\bm\psi_k\,\Delta\Xi_k.
\end{equation}
The subsequent macroscopic inner iterations update $\bm F_b^{\mathrm{out},m}$ and $\bm F_b^{\mathrm{in},m}$ in an incremental manner based on continuous-integration half-range moments, with the incoming part $\bm F_b^{\mathrm{in},m}$ closed by boundary conditions consistent with the corresponding kinetic ones.


\subsubsection{General update template in macroscopic inner iterations}\label{subsubsec:macro_bc_update}

Let $\bm W^{m}$ denote the macroscopic state at the $m$-th macroscopic inner iteration (within GSIS step $n\to n+1$, i.e. here we omit the superscript $n+1/2$). Since we don't update the discrete VDF during the macroscopic inner iteration, we estimate the change of the outgoing boundary flux from the change of $\bm W^{0}\to\bm W^{m}$ by assuming a Maxwellian distribution, and according to the definition \eqref{eq:Floc_plus_def} we get
\begin{equation}\label{eq:macro_Fout_update}
\bm F_{b}^{\mathrm{out},m} = \bm F_{b}^{\mathrm{out},m=0}+\mathbf P\!\left(\bm F_{\mathrm{loc}}^{+}(\bm W_{\mathrm{loc}}^{m})-\bm F_{\mathrm{loc}}^{+}(\bm W_{\mathrm{loc}}^{m=0})\right),  
\end{equation}
Note that here the flux increment is computed by the continuous-integration half-range moment.
Similarly, the change of the incoming part $\bm F_{b}^{\mathrm{in},m}$ is also evaluated by the continuous-integration moment, and according to the definition \eqref{eq:Fminus_hat_def} it reads
\begin{equation}\label{eq:macro_Fin_update}
\bm F_{b}^{\mathrm{in},m} = \bm F_{b}^{\mathrm{in},m=0}+ \mathbf P\!\left(\tilde \rho_w^{m}\widehat{\bm F}_{\mathrm{loc}}^{-}(\tilde T_w^{m}) - \tilde \rho_w^{m=0}\widehat{\bm F}_{\mathrm{loc}}^{-}(\tilde T_w^{m=0})\right),
\end{equation}
where $ \tilde \rho_w^{m},\tilde T_w^{m}$ depend on the boundary type and are calculated based on the outgoing flux $\bm F_{b}^{\mathrm{out},m}$, with the method described later. Then the total macroscopic boundary flux is
\begin{equation}\label{eq:macro_flux_total}
\bm F_b^{m}=\bm F_b^{\mathrm{out},m}+\bm F_b^{\mathrm{in},m}.
\end{equation}
Finally, after the macroscopic synthetic iterations, parameters $\rho _w^{n + 1},T _w^{n + 1}$ for the kinetic solver can be updated from the intermediate n + 1/2 step as:
\begin{equation}\label{eq:wall_update}
\begin{aligned}
&\rho _w^{n + 1} = \rho _w^{n + 1/2} + \tilde \rho _w^{m=M} - \tilde \rho _w^{m=0},\\
&T_w^{n + 1} = T_w^{n + 1/2} + \tilde T_w^{m=M} - \tilde T_w^{m=0}.
\end{aligned}
\end{equation}

The update scheme \eqref{eq:macro_Fout_update}-\eqref{eq:wall_update} allows the overall iterative algorithm converges correctly to a fixed point: When the macroscopic variable remains unchanged during the macroscopic synthetic iterations, i.e. $\bm W^{m}=\bm W^{m=0}$, the macroscopic boundary flux $\bm F_{b}^{m}$ is strictly equal to that obtained from the kinetic solver $\bm F_{b}^{m=0}$, and finally the predicted parameters $\rho _w^{n + 1},T _w^{n + 1}$ for the kinetic solver is strictly equal to $\rho _w^{n + 1/2},T _w^{n + 1/2}$.

\subsubsection{Isothermal wall}\label{subsubsec:macro_bc_iso}

For an isothermal diffuse wall, $\tilde T_w=T_w=T_0$ is prescribed and only $\tilde \rho_w$ is unknown. At each macroscopic iteration $m$, we update $\tilde \rho_w^{m}$ by enforcing the impermeability condition using the current predicted outgoing mass flux moment. Let $M^{+,m}$ denote the outgoing mass flux (the first component of the local outgoing flux moment), which is equivalently given by $M^{+,m}=(\bm F_{b}^{\mathrm{out},m})_{\rho}$. Combining the continuous incoming half-range mass moment $A(T_0)=\int_{v_n\le 0} v_n g_w(T_0)\mathrm d\bm v=-\sqrt{RT_0/2\pi}$ with the impermeability condition, we get
\begin{equation}\label{eq:rhow_iso_macro}
\tilde \rho_w^{m}=-\frac{M^{+,m}}{A(T_0)}=\frac{(\bm F_{b}^{\mathrm{out},m})_{\rho}}{\sqrt{RT_0/2\pi}}.
\end{equation}
The incoming flux $\bm F_{b}^{\mathrm{in},m}$ is then calculated by Eq.~\eqref{eq:macro_Fin_update}. Note that since both $\bm F_{b}^{\mathrm{in},m=0}$ and $\mathbf P\left( \tilde\rho_w^{m}\widehat{\bm F}_{\mathrm{loc}}^{-}(\tilde T_w^{m})\right)$ satisfy the impermeability constraint, it is easy to see that $\bm F_{b}^{\mathrm{in},m}$ obtained by  Eq.~\eqref{eq:macro_Fin_update} also strictly satisfies the impermeability constraint.

\subsubsection{Adiabatic wall and heat-flux wall}\label{subsubsec:macro_bc_adi_hf}

For adiabatic and heat-flux walls, both $\tilde T_w$ and $\tilde \rho_w$ are unknown. The incoming flux retains the diffuse-wall form, i.e. $\mathbf P\!\left(\tilde\rho_w^{m}\widehat{\bm F}_{\mathrm{loc}}^{-}(\tilde T_w^{m})\right)$.
The pair $(\tilde T_w^{m},\tilde\rho_w^{m})$ is updated by enforcing the impermeability and heat flux constraints as in Section~\ref{subsec:bc}, with the current predicted outgoing mass flux $M^{+,m}$ and energy flux $Q^{+,m}$ extracted from $\bm F_{b}^{\mathrm{out},m}$, the constraints read:
\begin{equation}\label{eq:wall_constraints_macro}
\begin{aligned}
& M^{+,m}+\tilde \rho_w^{m}A(\tilde T_w^{m})=0,\\
& Q^{+,m}+\tilde \rho_w^{m}B(\tilde T_w^{m})=-Q_n,
\end{aligned}
\end{equation}
where $Q_n=0$ for an adiabatic wall and $Q_n\neq 0$ for a heat-flux wall, and here the continuous-integration half-range moments is used:
\begin{equation}
    \begin{aligned}
       & A(T)=\int_{v_n\le 0} v_n g_w(T)\,\mathrm d\bm v=-\sqrt{\frac{RT}{2\pi}}, \\
& B(T)=\int_{v_n\le 0} \frac12 v^2 v_n g_w(T)\,\mathrm d\bm v=2RT\,A(T).
    \end{aligned}
\end{equation}
Solving Eq.~\eqref{eq:wall_constraints_macro} yields $(\tilde T_w^{m},\tilde \rho_w^{m})$, after which the incoming flux is evaluated by Eq.~\eqref{eq:macro_Fin_update}. Note that since both $\bm F_{b}^{\mathrm{in},m=0}$ and $\mathbf P\left( \tilde\rho_w^{m}\widehat{\bm F}_{\mathrm{loc}}^{-}(\tilde T_w^{m})\right)$ satisfy the impermeability and heat flux constraints, it is easy to see that $\bm F_{b}^{\mathrm{in},m}$ obtained by  Eq.~\eqref{eq:macro_Fin_update} also strictly satisfies the impermeability and heat flux constraints.

\subsubsection{Far-field boundary}\label{subsubsec:macro_bc_far}

At the far-field boundary, the incoming distribution on $\Xi^-$ is prescribed by the far-field Maxwellian $f_\infty(\rho_\infty,\bm u_\infty,T_\infty)$ and is independent of the macroscopic iterates. Therefore, the incoming contribution $\bm F_{b}^{\mathrm{in}}$ remains fixed during the macroscopic inner iterations:
\begin{equation}\label{eq:Fin_far_macro}
\bm F_{b}^{\mathrm{in},m}=\bm F_{b}^{\mathrm{in},0}.
\end{equation}
while the outgoing part is updated by Eq.~\eqref{eq:macro_Fout_update}. The total far-field flux then follows from Eq.~\eqref{eq:macro_flux_total}. This treatment preserves the kinetic open-boundary prescription while still allowing the macroscopic subsystem to update the outgoing contribution consistently.

In the macroscopic stage, the incoming wall-to-gas contribution is evaluated through the continuous half-space integrals \eqref{eq:Fminus_hat_def}--\eqref{eq:Fminus_hat_closed}, which are available in closed form and thus inexpensive and smooth. The wall-reflected parameters $(\tilde \rho_w,\tilde T_w)$ are updated so that the resulting macroscopic boundary flux satisfies the same physical constraints as the kinetic boundary condition, ensuring tight kinetic--macroscopic consistency throughout the GSIS iterations.

\section{Computational Results and Discussion}\label{sec:num_example}

This section assesses the accuracy, robustness, and convergence of the proposed boundary treatment within the GSIS framework for steady rarefied gas flows. Three representative problems are considered, including hypersonic flow through a 3D nozzle with fixed wall temperature, hypersonic flow past a 2D circular cylinder with adiabatic walls, and steady heat transfer in a 2D annulus with prescribed heat flux boundary condition, which is particularly sensitive to boundary heat-flux consistency. The working gas is argon with viscosity index $\omega = 0.81$, and convergence is declared when the volume-weighted relative error between two consecutive iterations satisfies $\varepsilon < 10^{-6}$ for $W\in\{\rho,\bm{u},T\}$:
\begin{equation}\label{convergence_critertion}
\varepsilon=\max \left(\sqrt{\frac{\sum_{i} \left ( W_i^{n}-W_i^{n-1} \right )^2 \varOmega _i }{\sum_{i} \left ( W_i^{n-1} \right )^2 \varOmega _i } }  \right).
\end{equation}
Note that all DSMC reference data in this study were obtained using the open-source software SPARTA with the variable-soft-sphere collision model~\cite{plimpton2015sparta}.

\subsection{Hypersonic flow through a 3D nozzle}\label{3DNozzle}

\begin{figure}[!t]
    \centering
    {\includegraphics[width=0.4\textwidth,clip = true]{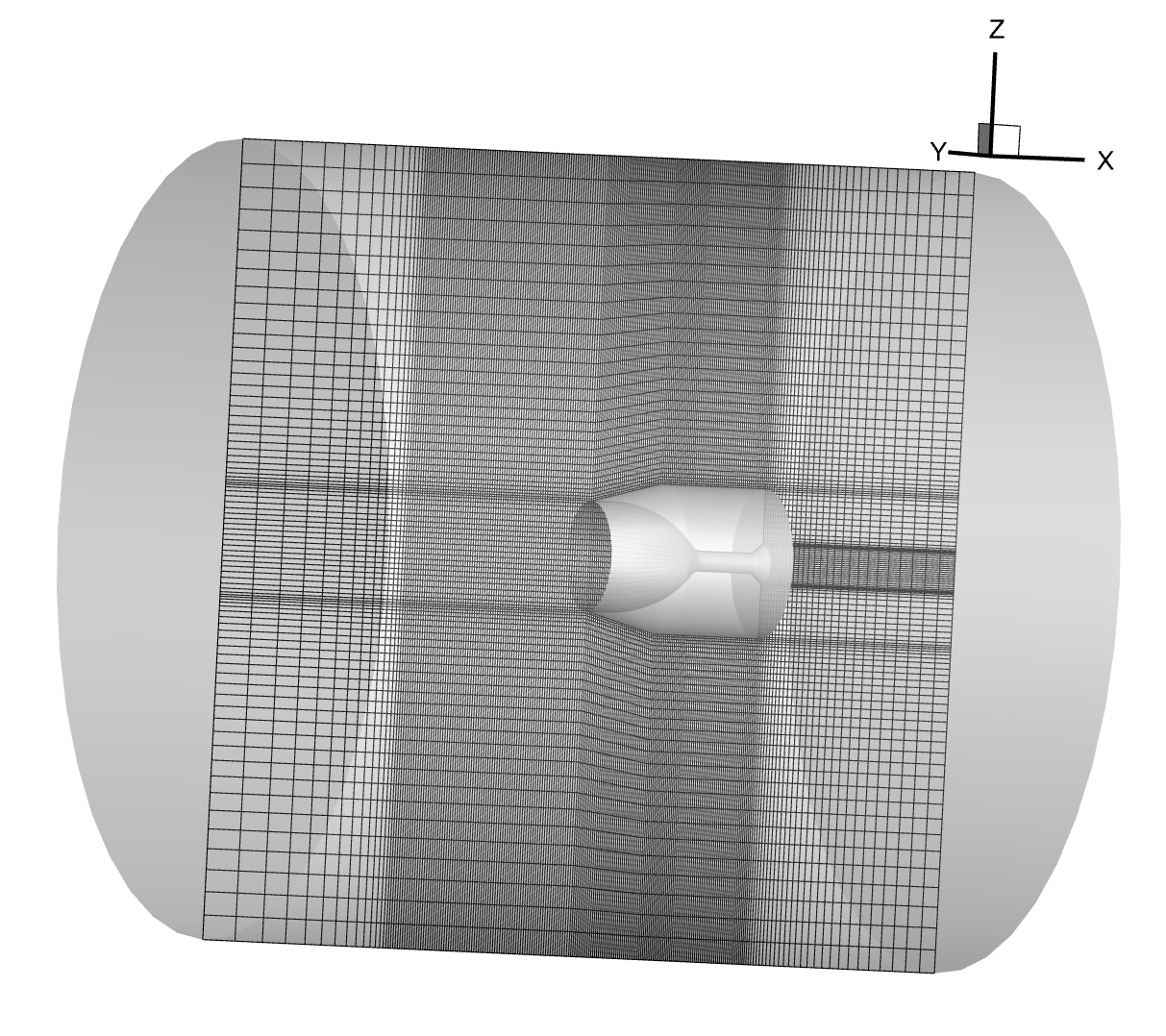}}
    \hspace{0.5cm}
    {\includegraphics[width=0.4\textwidth,clip = true]{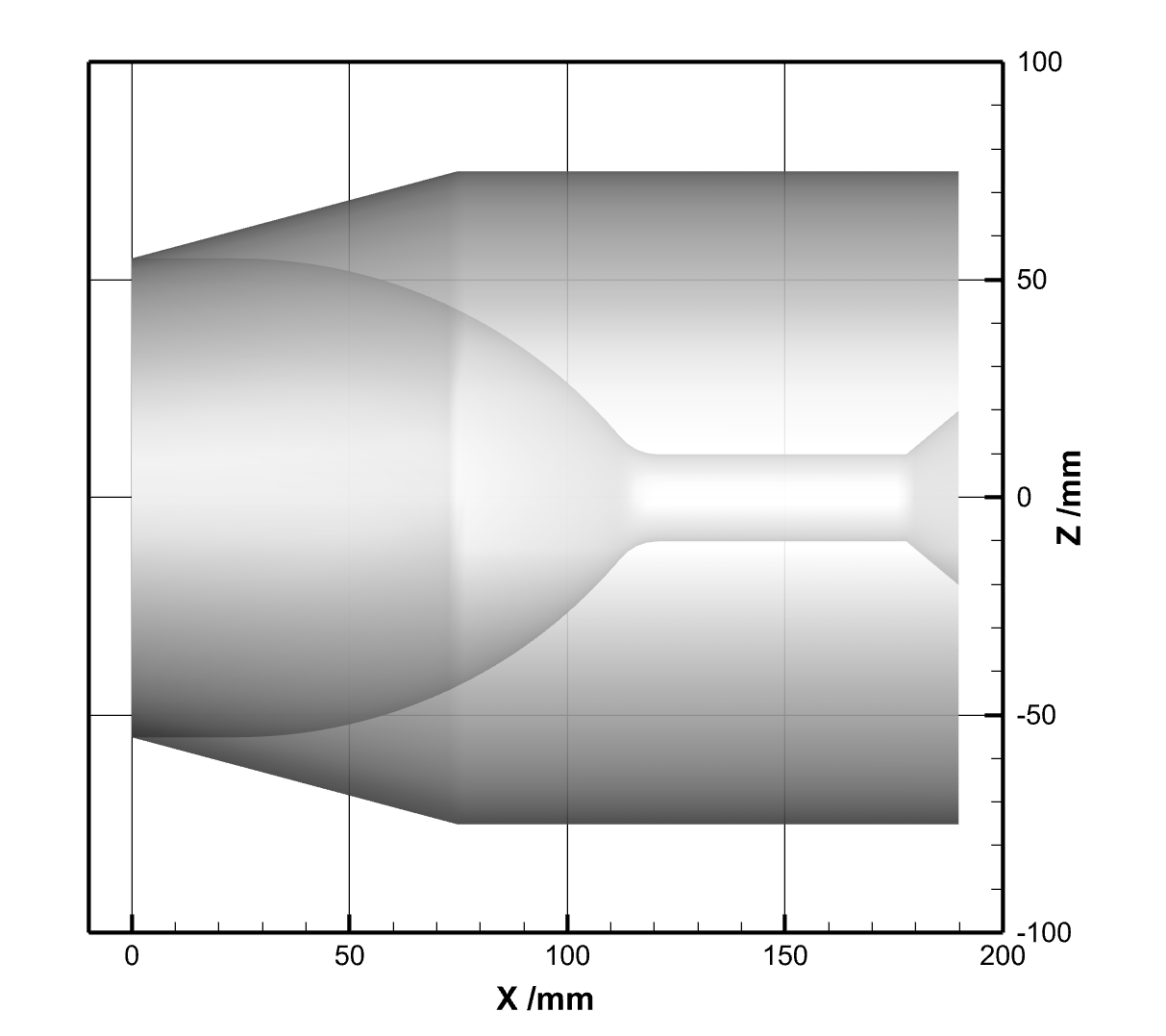}}
    \caption{
    3D computational mesh and nozzle geometry. Top: overall computational domain and hexahedral mesh with $335{,}740$ cells. Bottom: nozzle contour in the $x$--$z$ plane following the axisymmetric 2D geometry reported in~\cite{jin2024nozzle}. Only the geometric contour is adopted, while the present flow conditions and numerical setup are newly defined.
    }
    \label{fig:3DNozzle_Mesh}
\end{figure}

Consider a hypersonic internal-flow case through a 3D convergent--divergent nozzle.
The nozzle contour is taken from the axisymmetric 2D geometry in Ref.~\cite{jin2024nozzle} and is extended to a fully 3D configuration, see Figure~\ref{fig:3DNozzle_Mesh}.
The geometry consists of four parts, including an entrance section, a compression section, a throat section, and an exit section, with a  total nozzle length of $L_0=189.531~\mathrm{mm}$.
The inflow Mach number is set to $\mathrm{Ma}=5$ with the temperature $T_\infty=T_0=273.15~\mathrm{K}$, and the nozzle wall is maintained at an isothermal temperature $T_w=300~\mathrm{K}$.

\begin{figure*}[!t]
    \centering
    {\includegraphics[width=0.4\textwidth,clip = true]{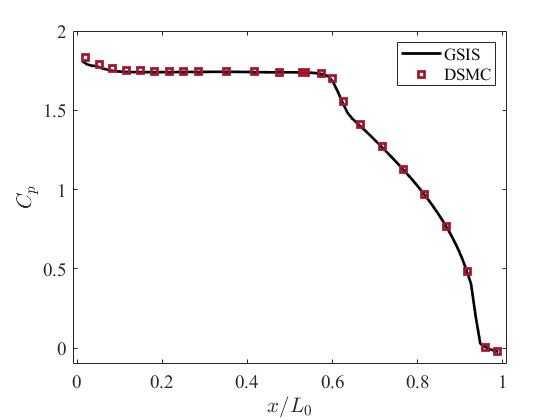}}
    {\includegraphics[width=0.4\textwidth,clip = true]{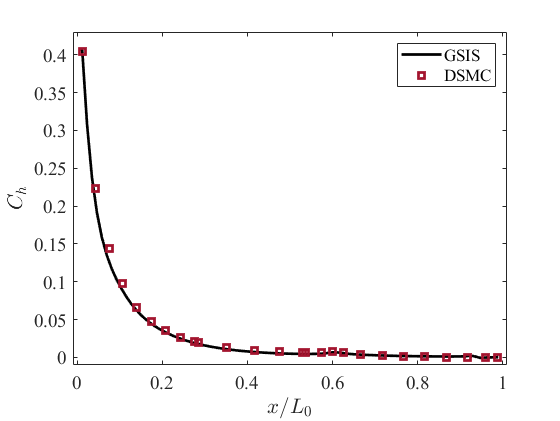}}
    \\
    {\includegraphics[width=0.4\textwidth,clip = true]{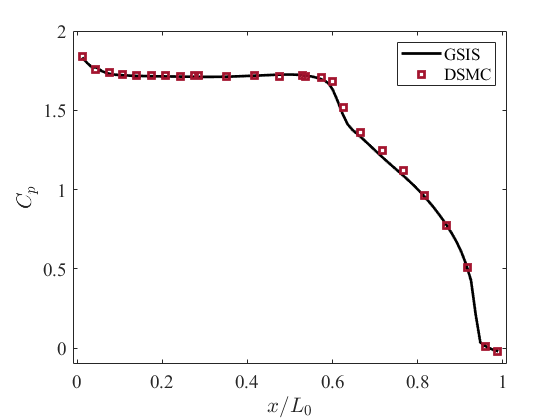}}
    {\includegraphics[width=0.4\textwidth,clip = true]{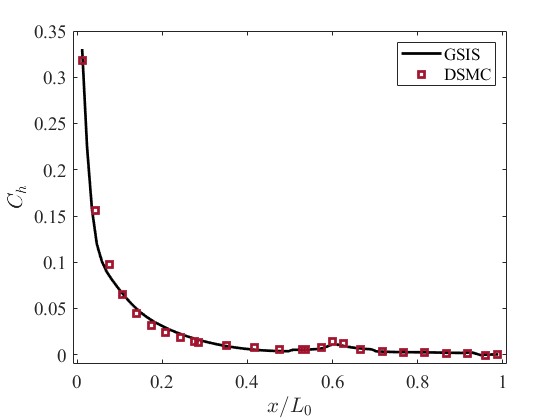}}
    \caption{Comparisons of \(C_p\) and \(C_h\) along the nozzle inner wall as functions of \(x/L_0\) between DSMC and GSIS at \(\mathrm{Kn}=0.5276\) (top) and \(\mathrm{Kn}=0.2638\) (bottom) with \(\mathrm{Ma}=5\).}
\label{fig:3DNozzle_Ma5_DSMC_GSIS_cmp}
\end{figure*}

To assess rarefaction effects, two inflow number densities are considered, $n_{\infty}=1.6780\times10^{19}~\mathrm{m^{-3}}$ and $3.3559\times10^{19}~\mathrm{m^{-3}}$, which correspond to $\mathrm{Kn}=0.5276$ and $0.2638$, respectively, based on the reference length $L_0$.
The molecular velocity space is truncated to $[-10,14]\sqrt{RT_0}\times[-12,12]\sqrt{RT_0}\times[-12,12]\sqrt{RT_0}$ and discretized by a  $28\times28\times28$ uniform grid.

Figure~\ref{fig:3DNozzle_Ma5_DSMC_GSIS_cmp} compares the distributions of the drag coefficient \( C_p \) and the heat transfer coefficient \( C_h \) predicted by GSIS with those obtained using DSMC.
These dimensionless coefficients are defined as:
\begin{equation}
    \begin{aligned}
        C_{p} = \frac{p_{w} - p_{\infty}}{\frac{1}{2}\rho_{\infty}u_{\infty}^{2}}, \quad
        C_{h} = \frac{q_{w}}{ \frac{1}{2}\rho_{\infty}u_{\infty}^{3} }.
    \end{aligned}
\end{equation}
where \(p_w\) and \(q_w\) denote the wall-normal momentum flux and energy flux from the gas to the wall:
\begin{equation}
    \begin{aligned}
        p_w &= \sum_{\bm{v}_k\in\Xi} m (\bm{v}_k\!\cdot\!\bm{n}_w)\ (\bm{v}_k\!\cdot\!\bm{n}_w)\,f_{w,k}\,\Delta\Xi_k, \\
        q_w &= \sum_{\bm{v}_k\in\Xi} \frac{1}{2} m v_k^2 \ (\bm{v}_k\!\cdot\!\bm{n}_w)\,f_{w,k}\,\Delta\Xi_k.
    \end{aligned}
\end{equation}
with \(\bm{n}_w\) being the outward unit normal pointing from the gas toward the wall.
Both surface quantities show excellent agreement, demonstrating that the current boundary setup and coupled GSIS iteration can reliably predict surface aerothermodynamic responses on a complex 3D nozzle geometry across the tested rarefaction levels. 
Notably, with varying Knudsen number, the wall heat flux (hence $C_h$) demonstrates greater sensitivity, attributable to its nature as a higher-order non-equilibrium transport quantity under pronounced Knudsen layer and temperature jump influences; conversely, the wall-normal stress (hence $C_p$) remains largely governed by the compressible pressure field and impermeability constraint, making it less responsive to rarefaction effects within the considered parameter range.

Note that the two-dimensional axisymmetric DSMC simulations employed locally refined meshes with 597,594 and 907,461 cells and 50 particles per cell for the $\mathrm{Kn}=0.5726$ and $\mathrm{Kn}=0.2638$ cases, respectively. Owing to its AP property~\cite{su2020fast}, GSIS can employ cell sizes much larger than the molecular mean free path. Consequently, only 335,740 cells are sufficient for the three-dimensional simulations.
On an Intel Xeon Silver 4208 CPU@2.10 GHz platform using 192 cores, the GSIS solver required only 0.51 and 0.45 wall-clock hours for the $\mathrm{Kn}=0.5726$ and $\mathrm{Kn}=0.2638$ cases, respectively. By comparison, CIS simulations on the same platform required 25.1 and 33.7 hours. These results validate the proposed boundary formulation and highlight its substantial computational efficiency.


\begin{figure}[t]
\centering
    {\includegraphics[width=0.48\textwidth,trim={20 0 20 90},clip = true]{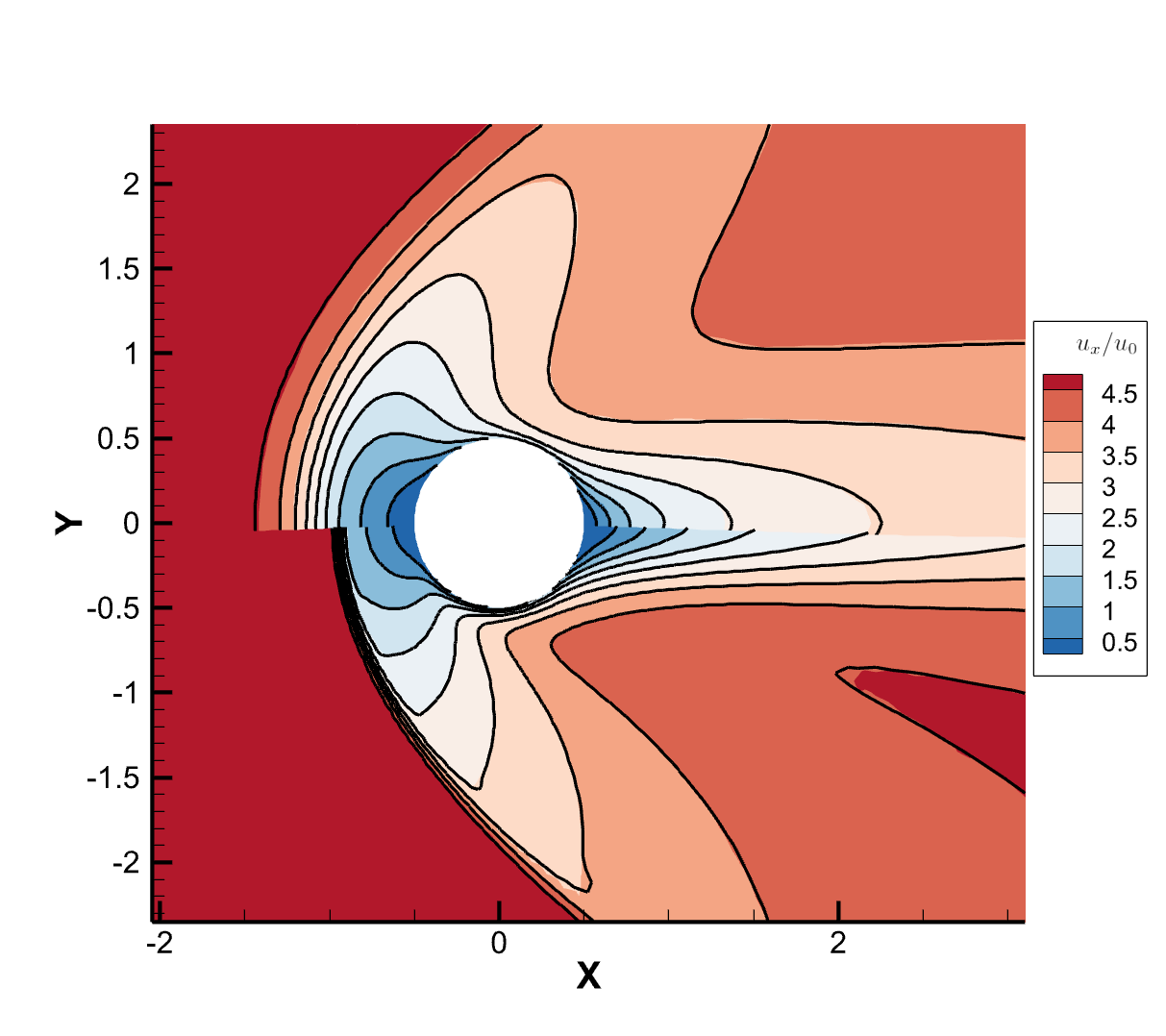}}
    {\includegraphics[width=0.48\textwidth,trim={20 0 20 90},clip = true]{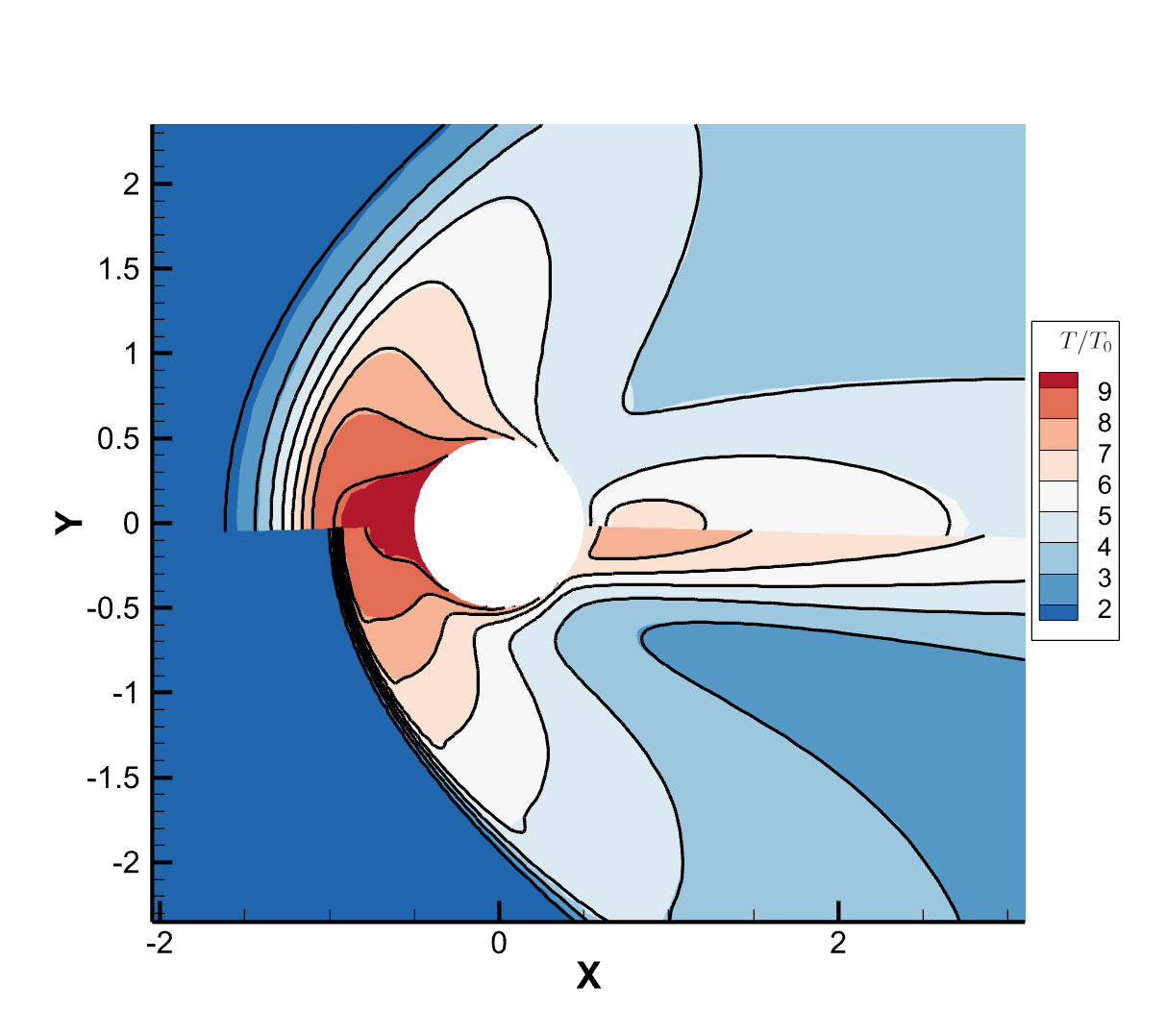}}
\caption{Comparisons of velocity and temperature between DSMC (contours) and GSIS (black lines) with $\text{Ma} = 5$. In each figure, the top and bottom half spatial regions show the results of $\text{Kn} = 0.1$ and 0.01, respectively.}
\label{fig:2DCylinder21120_Ma5_DSMC_GSIS_cmp}
\end{figure}

\subsection{Hypersonic flow around an adiabatic circular cylinder}\label{2DCylinder}

We next examine the hypersonic flow at $\mathrm{Ma}=5$ passing over a circular cylinder. The inflow temperature is set to $T_0$. The cylinder surface is adiabatic, and we assume fully diffuse gas-surface interactions; therefore, the wall temperature has to be determined during the numerical simulation. The reference length is the cylinder diameter, denoted as $L_0 = 1~\mathrm{m}$. Our computational domain is a circle with a radius of $5\ L_0$, discretized using $240\times 88$ quadrilateral cells. The first layer next to the wall has a thickness of $0.005\ L_0$. Simulations are performed for $\mathrm{Kn}=0.1$ and $0.01$.

This configuration contains both near-equilibrium regions (e.g., upstream of the shock) and strongly non-equilibrium regions (e.g., the shock layer and the wall-adjacent zone), which provides a stringent test for the coupled kinetic--synthetic iteration and the associated boundary treatment. For GSIS, the truncated velocity space $[-10\sqrt{RT_0},\, 14\sqrt{RT_0}]\times[-12\sqrt{RT_0},\, 12\sqrt{RT_0}]$ is discretized uniformly into $60\times 60$ cells, and the same discretization is used for both $\mathrm{Kn}=0.1$ and $0.01$.



Figure~\ref{fig:2DCylinder21120_Ma5_DSMC_GSIS_cmp} compares the GSIS and DSMC results for the streamwise velocity and temperature, where $u_0=\sqrt{(5/3) R T_0}$ is used for normalization. Overall, good agreement is observed for both Knudsen numbers, indicating that the present boundary setup and the coupled GSIS iteration capture the key non-equilibrium features in the shock layer and the wall-adjacent region. A noticeable discrepancy appears in the temperature field at $\mathrm{Kn}=0.1$, where the Shakhov model exhibits a slightly more upstream temperature rise ahead of the bow shock (i.e., an elongated upstream temperature tail) compared with DSMC. This behavior is consistent with the known limitations of the single-relaxation kinetic models in strong shock-structure computations, where the simplified relaxation collision term cannot fully reproduce the Boltzmann collision dynamics of higher-order moments and may therefore broaden the temperature profile upstream~\cite{liu2014investigation}. As the Knudsen number decreases (e.g., $\mathrm{Kn}=0.01$), intermolecular collisions become dominant and the flow approaches local equilibrium, so the upstream preheating/early-rise artifact is mitigated and the temperature alignment with DSMC is improved.

This test case validates that the proposed boundary formulation and its numerical implementation are reasonable and effective.  
At $\mathrm{Kn}=0.1$ and $0.01$, GSIS required 0.09 and 0.12 wall-clock hours on 8 cores, while DSMC needed 0.15 and 2.90 hours on 400 cores (33\,808 and 2\,010\,616 cells, 100 particles per cell), underscoring the pronounced computational advantage of GSIS--especially at lower Knudsen numbers--while maintaining accuracy.


\begin{figure*}[!t]
    \centering
    \subfigure[\label{fig:HeatFlux_Kn0d01_gsis_dvm_cmp_T}]{\includegraphics[width=0.32\textwidth,trim={20 20 20 40}, clip = true]{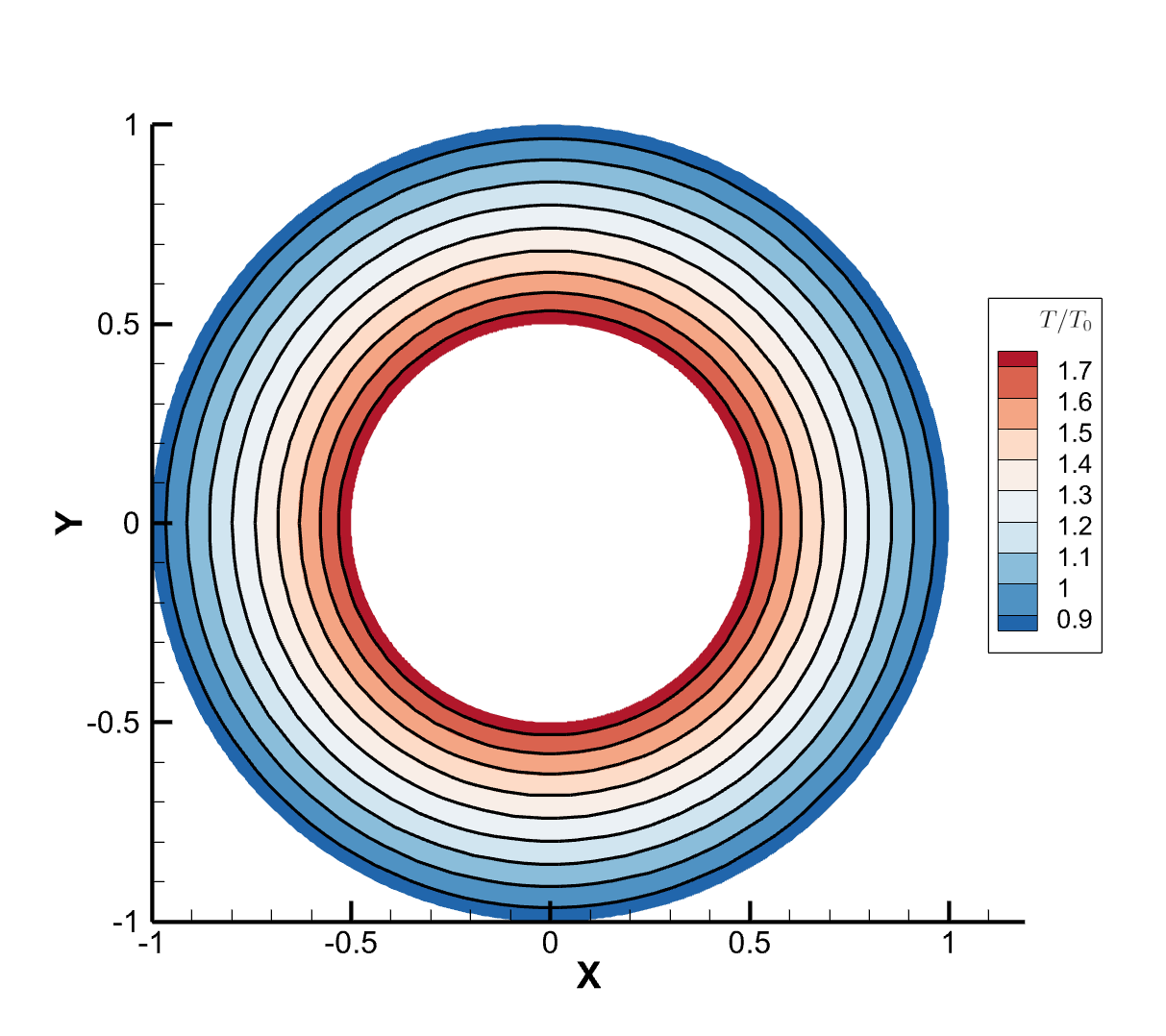}}
    \subfigure[\label{fig:HeatFlux_Kn0d01_gsis_dvm_cmp_qx}]{\includegraphics[width=0.32\textwidth, trim={20 20 20 40}, clip = true]{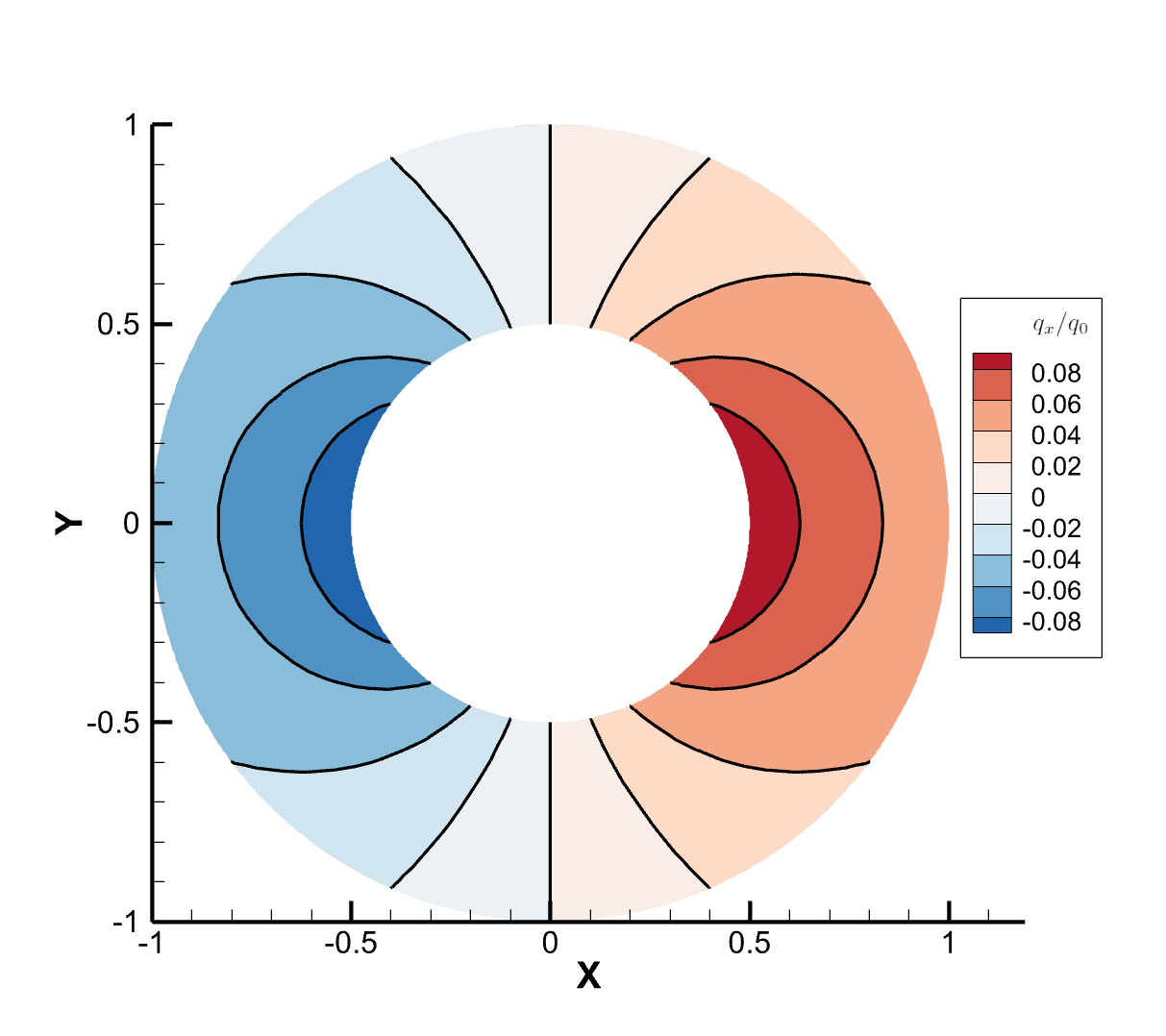}}
    \subfigure[\label{fig:HeatFlux_Kn0d01_gsis_dvm_Steps_Qout}]{\includegraphics[width=0.34\textwidth,clip = true]{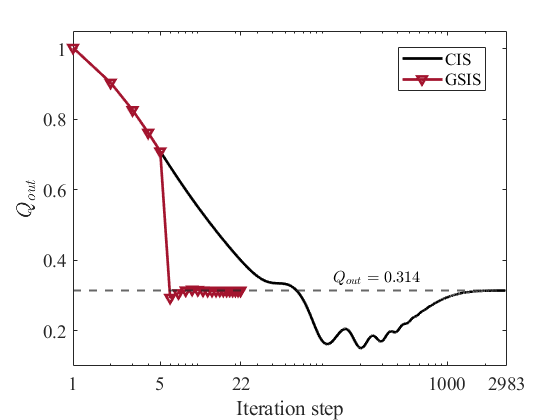}}
    \caption{For the annular configuration with $\text{Kn}=0.01$, comparisons between the CIS and the GSIS method are presented for the (a) temperature, (b) heat flux, and (c) the convergence history of the net heat flux at the outer isothermal wall.
    Note that the first five steps in the GSIS method evolve the initial field using the CIS update.
}
\label{fig:HeatFlux_Kn0d01_CIS_GSIS_cmp}
\end{figure*}

\subsection{Heat transfer in an annulus with an inner heat-flux wall}\label{Heat-Flux}

We finally examine steady heat transfer in a 2D annulus at $\mathrm{Kn}=0.01$ with reference length $L_0 = 1\mathrm{m}$, aiming to test the robustness of the boundary treatment for thermal problems and assess the consistency of the wall heat-flux evaluation. As shown in Fig.~\ref{fig:HeatFlux_Kn0d01_gsis_dvm_cmp_T}, the inner cylinder is a prescribed heat-flux wall with $Q_n=0.1\,Q_0$ (the reference heat-flux is defined as $Q_0 = n_0k_BT_0\sqrt{RT_0}$) injecting heat into the gas, and the outer cylinder is an isothermal wall maintained at $T_w = 0.8\,T_0$. The spatial mesh, which has been verified to be grid-independent, consists of 2,520 cells with the first-layer height set to $0.005\,L_0$, and the truncated velocity space $[-8\sqrt{RT_0},\, 8\sqrt{RT_0}]^2$ is uniformly discretized into a $60\times 60$ grid. 

The outward unit normal is defined to point from the gas to the wall, and the net heat transfer through the outer isothermal boundary is quantified by the outward heat flux:
\begin{equation}\label{eq:Qout_def}
Q_{out}=\sum_{j \in \Gamma_{o}} \sum_{\bm{v}_k\in\Xi} \frac{1}{2}m v_k^2\,(\bm{v}_k\!\cdot\!\bm{n}_j)\,f_{j, k}\,\Delta\Xi_k S_j.
\end{equation}
Here, $\Gamma_{o}$ denotes the outer isothermal wall boundary, the interface distribution function $f_{j,k}$ is obtained by an upwind reconstruction of the cell-centered distribution function.

Figure~\ref{fig:HeatFlux_Kn0d01_gsis_dvm_cmp_T} compares the temperature field obtained by CIS and GSIS, and Fig.~\ref{fig:HeatFlux_Kn0d01_gsis_dvm_cmp_qx} shows the distribution of the heat-flux component, where both methods exhibit close agreement in the bulk region and near the boundaries. The net heat input through the inner heat-flux wall is $Q_{in}=0.314\,Q_0$ (note that $Q_n$ is a local prescribed normal heat flux defined on a single boundary face), which is obtained by summing the prescribed heat-flux contribution over the inner cylindrical boundary, and energy conservation at steady state requires $Q_{out}=Q_{in}$. Therefore, the evolution of $Q_{out}$ provides a direct and sensitive indicator of the convergence of the entire flow field, and it is particularly suitable for monitoring steady thermal problems with mixed heat-flux/isothermal boundaries.

Figure~\ref{fig:HeatFlux_Kn0d01_gsis_dvm_Steps_Qout} reports the iteration history of $Q_{out}$ defined in Eq.~\eqref{eq:Qout_def}. In GSIS, the first five steps are used to evolve the initial field using the CIS update; the first application of the synthetic equation in GSIS iteration already transfers the heat-flux information imposed at the inner cylindrical boundary to the outer cylindrical boundary, indicating an immediate global coupling of the thermal field. After only a few subsequent adjustment iterations, the entire solution reaches convergence. In contrast, the CIS gradually propagates boundary information outward through successive streaming iterations, leading to much slower convergence for this setup and even overshooting. Consequently, GSIS demonstrates a significant advantage over CIS as it reduces the iteration numbers by nearly two orders of magnitude.

\section{Conclusions}\label{sec:conclusion}

This work presents an efficient macroscopic boundary treatment for isothermal, adiabatic, and prescribed heat-flux walls within the GSIS. The formulation updates boundary fluxes in each macroscopic inner iteration via explicit half-space splitting: the outgoing contribution is obtained from equilibrium half-range moments of the evolving macroscopic state, while the incoming contribution is evaluated analytically from closed-form Maxwellian half-range moments. The wall parameters $(T_w,\rho_w)$ are determined by enforcing the same fully discrete impermeability and energy-flux constraints as in the kinetic diffuse-wall condition, thereby maintaining discrete kinetic--macroscopic consistency and avoiding near-wall convergence degradation.

The method is assessed on three representative problems: hypersonic rarefied flow in a 3D nozzle with isothermal walls, hypersonic flow around a 2D adiabatic circular cylinder, and steady heat transfer in a 2D annulus with an inner heat-flux wall and an outer isothermal wall. Across these cases, GSIS shows good agreement with DSMC for the nozzle and cylinder cases, and with conventional CIS for the annulus heat-transfer case, while delivering substantial efficiency gains; for example, it reduces the wall-clock time by about 50$\times$ relative to conventional CIS in the nozzle case and cuts the iteration count by nearly two orders of magnitude in the 2D steady heat-transfer case. Overall, the proposed boundary formulation offers a practical and robust route for GSIS simulations involving prescribed thermal wall conditions. Future work will extend the approach to advanced gas models and gas--surface interaction models incorporating finite thermal accommodation coefficients, and further to chemically reacting flows


\section*{Acknowledgments} 
This work is supported by the National Natural Science Foundation of China (125B2052, 12402388). The authors acknowledge the computing resources from the Center for Computational Science and Engineering at the Southern University of Science and Technology.


\bibliographystyle{elsarticle-num}

\bibliography{ref}

@article{Bhatnagar1954,
  title={{A model for collision processes in gases. I. Small amplitude processes in charged and neutral one-component systems}},
  author={Bhatnagar, Prabhu Lal and Gross, Eugene P and Krook, Max},
  journal={Physical Review},
  volume={94},
  pages={511},
  year={1954},
}

@article{Fei2023JCP,
  title={{A time-relaxed Monte Carlo method preserving the Navier-Stokes asymptotics}},
  author={Fei, Fei},
  journal={Journal of Computational Physics},
  volume={486},
  pages={112128},
  year={2023},
  publisher={Elsevier}}

@article{Gorji2014JCP,
  title={An efficient particle {Fokker--Planck} algorithm for rarefied gas flows},
  author={Gorji, M Hossein and Jenny, Patrick},
  journal={Journal of Computational Physics},
  volume={262},
  pages={325--343},
  year={2014},
  publisher={Elsevier}}

@article{LiuZhu2020JCP,
	author = {C. Liu and Y. J. Zhu and K. Xu},
	journal = {Journal of Computational Physics},
	pages = {108977},
	title = {{Unified gas-kinetic wave-particle methods I: Continuum and rarefied gas flow}},
	volume = {401},
	year = {2020}}

@article{Dimarco2018JCP,
	author = {G. Dimarco and R. Loub{\`e}re and J. Narski and T. Rey},
	journal = {Journal of Computational Physics},
	pages = {46-81},
	title = {{An efficient numerical method for solving the Boltzmann equation in multidimensions}},
	volume = {353},
	year = {2018}}

@article{Zeng2023CaF,
  title={General synthetic iterative scheme for polyatomic rarefied gas flows},
  author={Zeng, J. N. and Yuan, R. F. and Zhang, Y. B. and Li, Q and Wu, L},
  journal={Computers \& Fluids},
  volume={265},
  pages={105998},
  year={2023},
  publisher={Elsevier}}

@book{bird1994molecular,
  title={Molecular gas dynamics and the direct simulation of gas flows},
  author={Bird, Graeme A},
  year={1994},
  publisher={Oxford university press}}

@article{su2020can,
	author = {Su, Wei and Zhu, L. H. and Wang, Peng and Zhang, Y. H. and Wu, Lei},
	journal = {Journal of Computational Physics},
	pages = {109245},
	title = {Can we find steady-state solutions to multiscale rarefied gas flows within dozens of iterations?},
	volume = {407},
	year = {2020}}

@article{su2020fast,
	author = {Su, Wei and Zhu, L. H. and Wu, Lei},
	journal = {SIAM Journal on Scientific Computing},
	pages = {B1517--B1544},
	publisher = {SIAM},
	title = {Fast convergence and asymptotic preserving of the General Synthetic Iterative Scheme},
	volume = {42},
	year = {2020}}

@article{zhu2021general,
  title={General synthetic iterative scheme for nonlinear gas kinetic simulation of multi-scale rarefied gas flows},
  author={Zhu, L. H. and Pi, X. C. and Su, Wei and Li, Z. H. and Zhang, Y. H. and Wu, Lei},
  journal={Journal of Computational Physics},
  volume={430},
  pages={110091},
  year={2021},
  publisher={Elsevier}}

@article{guo2013discrete,
	author = {Guo, Z. L. and Xu, Kun and Wang, R. J.},
	journal = {Physical Review E},
	pages = {033305},
	title = {Discrete unified gas kinetic scheme for all {Knudsen} number flows: Low-speed isothermal case},
	volume = {88},
	year = {2013}}

@book{chapman1990mathematical,
	author = {Chapman, Sydney and Cowling, Thomas George},
	publisher = {Cambridge university press},
	title = {The mathematical theory of non-uniform gases: an account of the kinetic theory of viscosity, thermal conduction and diffusion in gases},
	year = {1990}}

@article{zhu2016implicit,
	author = {Zhu, Y. J. and Zhong, C. W. and Xu, Kun},
	journal = {Journal of Computational Physics},
	pages = {16--38},
	publisher = {Elsevier},
	title = {Implicit unified gas-kinetic scheme for steady state solutions in all flow regimes},
	volume = {315},
	year = {2016}}

@article{xu2010unified,
	author = {Xu, Kun and Huang, Juan-Chen},
	journal = {Journal of Computational Physics},
	pages = {7747--7764},
	publisher = {Elsevier},
	title = {A unified gas-kinetic scheme for continuum and rarefied flows},
	volume = {229},
	year = {2010}}

@article{liu2024further,
  title={Further acceleration of multiscale simulation of rarefied gas flow via a generalized boundary treatment},
  author={Liu, W and Zhang, Y. B. and Zeng, J. N. and Wu, L},
  journal={Journal of Computational Physics},
  volume={503},
  pages={112830},
  year={2024},
  publisher={Elsevier}}

@article{plimpton2015sparta,
	author = {Plimpton, SJ and Gallis, MA},
	journal = {Sandia National Laboratories, USA, see http://sparta. sandia. gov},
	title = {SPARTA direct simulation Monte Carlo (DSMC) simulator},
	year = {2015}}

@article{zhang2024efficient,
  title={Efficient parallel solver for rarefied gas flow using {GSIS}},
  author={Zhang, Y. B. and Zeng, J. N. and Yuan, R. F. and Liu, W and Li, Q and Wu, Lei},
  journal={Computers \& Fluids},
  volume={281},
  pages={106374},
  year={2024},
  publisher={Elsevier}}

@article{sun2004hybrid,
	author = {Sun, Quanhua and Boyd, Iain D and Candler, Graham V},
	journal = {Journal of Computational Physics},
	number = {1},
	pages = {256--277},
	publisher = {Elsevier},
	title = {A hybrid continuum/particle approach for modeling subsonic, rarefied gas flows},
	volume = {194},
	year = {2004}}

@article{AkhlaghiRoohi2016DSMCHeatFlux,
  author       = {Hassan Akhlaghi and Ehsan Roohi},
  title        = {A novel algorithm for implementing a specified wall heat flux in {DSMC}: Application to micro/nano flows and hypersonic flows},
  journal      = {Computers \& Fluids},
  volume       = {127},
  pages        = {78--101},
  year         = {2016},
}

@article{Fei2020UnifiedBGK,
  title   = {A unified stochastic particle {B}hatnagar-{G}ross-{K}rook method for multiscale gas flows},
  author  = {Fei, F. and Zhang, J. and Li, J. and Liu, Z. H.},
  journal = {Journal of Computational Physics},
  volume  = {400},
  pages   = {108972},
  year    = {2020}
}

@article{liu2014investigation,
  title={Investigation of the kinetic model equations},
  author={Liu, S and Zhong, C. W.},
  journal={Physical Review E},
  volume={89},
  number={3},
  pages={033306},
  year={2014},
  publisher={APS}
}

@article{torrilhon2008r13bc,
  title={Boundary conditions for regularized 13-moment-equations for micro-channel-flows},
  author={Torrilhon, Manuel and Struchtrup, Henning},
  journal={Journal of Computational Physics},
  volume={227},
  number={3},
  pages={1982--2011},
  year={2008},
  publisher={Elsevier}
}

@article{thatcher2008g13bc,
  title={Boundary conditions for {Grad's} 13 moment equations},
  author={Thatcher, Toby and Zheng, Y and Struchtrup, H},
  journal={Progress in Computational Fluid Dynamics, an International Journal},
  volume={8},
  number={1-4},
  pages={69--83},
  year={2008},
  publisher={Inderscience Publishers}
}

@article{shakhov1968,
  title={Approximate kinetic equations in rarefied gas theory},
  author={Shakhov, EM},
  journal={Fluid Dynamics},
  volume={3},
  number={1},
  pages={112--115},
  year={1968},
  publisher={Springer}
}

@article{shima2011slau,
  title={Parameter-free simple low-dissipation {AUSM-family} scheme for all speeds},
  author={Shima, Eiji and Kitamura, Keiichi},
  journal={AIAA journal},
  volume={49},
  number={8},
  pages={1693--1709},
  year={2011}
}

@Book{Aristov2001Direct,
  author    = {V.V. Aristov},
  publisher = {Springer},
  title     = {{Direct Methods for Solving the Boltzmann Equation and Study of Nonequilibrium Flows}},
  year      = {2001},
}

@article{jin2024nozzle,
  title={Numerical and experimental investigation of rarefied hypersonic flow in a nozzle},
  author={Jin, X. H. and Su, P. H. and Chen, Z. and Cheng, X. L. and Wang, Qiang and Wang, Bing},
  journal={Physics of Fluids},
  volume={36},
  number={11},
  year={2024},
  publisher={AIP Publishing}
}

@article{Luo2026DIG,
  title={{Enhancing DSMC simulations of rarefied gas mixtures using a fast-converging and asymptotic-preserving scheme}},
  author={L. Y. Luo and J. N. Zeng and Y. B. Zhang and W. Li and Q. Li and L. Wu},
  journal={Computer Methods in Applied Mechanics and Engineering},
  volume={449},
  pages={118508},
  year={2026},
}

@article{Zeng2025GSIS,
  title={General synthetic iterative scheme for rarefied gas mixture flows},
  author={J. N. Zeng and Q. Li and L. Wu},
  journal={Journal of Computational Physics},
  volume={519},
  pages={113420},
  year={2025},
}

@article{Luo2024DIG,
  title={{Multiscale simulation of rarefied gas dynamics via direct intermittent GSIS-DSMC coupling}},
  author={L. Y. Luo and L. Wu},
  journal={Advances in Aerodynamics},
  volume={6},
  pages={22},
  year={2024},
}

@article{Zeng2026GSIS,
  title={Accelerated simulation of multiscale gas-radiation coupling flows via a general synthetic iterative scheme},
  author={J. N. Zeng and Q. Li and Y. B. Zhang and W. Su and L. Wu},
  journal={arXiv},
  year={2026},
}

@article{mieussens2000,
  title={Discrete-velocity models and numerical schemes for the {Boltzmann-BGK} equation in plane and axisymmetric geometries},
  author={Mieussens, Luc},
  journal={Journal of Computational Physics},
  volume={162},
  pages={429--466},
  year={2000}
}

@article{klar1999,
  title={An asymptotic preserving numerical scheme for kinetic equations in the low Mach number limit},
  author={Klar, Axel},
  journal={SIAM journal on numerical analysis},
  volume={36},
  number={5},
  pages={1507--1527},
  year={1999},
  publisher={SIAM}
}

@article{dimarcoPareschi2013,
  title={Asymptotic preserving implicit-explicit {Runge--Kutta} methods for nonlinear kinetic equations},
  author={Dimarco, Giacomo and Pareschi, Lorenzo},
  journal={SIAM Journal on Numerical Analysis},
  volume={51},
  pages={1064--1087},
  year={2013},
}

@article{Khan2021ConstHeatFlux,
  title   = {Experimentally-Benchmarked kinetic simulations of heat transfer through rarefied gas with constant heat flux at the boundary},
  author  = {Khan, M. Adnan and Jobic, Yann and Graur, Irina and Hadj-Nacer, Mustafa and Zampella, Cody and Greiner, Miles},
  journal = {International Journal of Heat and Mass Transfer},
  volume  = {176},
  pages   = {121378},
  year    = {2021},
}

@article{Meng2015HeatFluxBC,
  title   = {Numerical Simulation of Rarefied Gas Flows with Specified Heat Flux Boundary Conditions},
  author  = {Meng, Jian Ping and Zhang, Yong Hao and Reese, Jason M.},
  journal = {Communications in Computational Physics},
  volume  = {17},
  pages   = {1185--1200},
  year    = {2015},
}

@article{akhlaghi2012new,
  title={A new iterative wall heat flux specifying technique in {DSMC for heating/cooling simulations of MEMS/NEMS}},
  author={Akhlaghi, Hassan and Roohi, Ehsan and Stefanov, Stefan},
  journal={International Journal of Thermal Sciences},
  volume={59},
  pages={111--125},
  year={2012},
  publisher={Elsevier}
}

@article{wang2008heat,
  title={Heat-flux-specified boundary treatment for gas flow and heat transfer in microchannel using direct simulation Monte Carlo method},
  author={Wang, Qiu Wang and Yan, Xiao Hong and He, Qun Wu},
  journal={International Journal for Numerical Methods in Engineering},
  volume={74},
  pages={1109--1127},
  year={2008},
}

\end{document}